\documentclass[aps,prl,twocolumn,amsmath,amssymb]{revtex4-1}

\usepackage{epsfig}
\usepackage{graphicx}
\usepackage{amsfonts}
\usepackage{amssymb}
\usepackage{amsmath}
\usepackage{amsfonts}
\usepackage{bm}

\newcommand{\ket}[1]{| #1 \rangle}

\renewcommand{\vec}[1]{{\mathbf #1}}

\bibpunct{(}{)}{,}{s}{,}{,}

\def\be{\begin{equation}}
\def\ee{\end{equation}}
\def\bea{\begin{eqnarray}}
\def\eea{\end{eqnarray}}
\begin{document}

\title{Modular Anomalies in $(2+1)$ and $(3+1)$-D Edge Theories}

\author{Moon Jip Park$^{1,2}$}
\author{Chen Fang$^{1,2,3}$}
\author{B. Andrei Bernevig$^{4}$}
\author{Matthew J. Gilbert$^{2,5}$}
\affiliation{$^1$ Department of Physics, University of Illinois, Urbana, IL 61801}
\affiliation{$^2$ Micro and Nanotechnology Laboratory, University of Illinois, Urbana, IL 61801}
\affiliation{$^3$ Department of Physics, Massachusetts Institute of Technology, Cambridge, MA 02139}
\affiliation{$^4$ Department of Physics, Princeton University, Princeton, NJ 08544, USA}
\affiliation{$^5$ Department of Electrical and Computer Engineering, University of Illinois, Urbana, IL 61801}

\date{\today}

\begin{abstract}
The classification of topological phases of matter in the presence of interactions is an area of intense interest. One possible means of classification is via studying the partition function under modular transforms, as the presence of an anomalous phase arising in the edge theory of a D-dimensional system under modular transformation, or modular anomaly, signals the presence of a $(D+1)$-D non-trivial bulk. In this work, we discuss the modular transformations of conformal field theories along a $(2+1)$-D and a $(3+1)$-D edge. Using both analytical and numerical methods, we show that chiral complex free fermions in $(2+1)$-D and $(3+1)$-D are modular invariant. However, we show in $(3+1)$-D that when the edge theory is coupled to a background $U(1)$ gauge field this results in the presence of a modular anomaly that is the manifestation of a quantum Hall effect in a $(4+1)$-D bulk. Using the modular anomaly, we find that the edge theory of $(4+1)$-D insulator with spacetime inversion symmetry$(P*T)$ and fermion number parity symmetry for each spin becomes modular invariant when $8$ copies of the edges exist.
\end{abstract}
\maketitle

The quantum Hall effect\cite{klitzing1980} has been an intense area of research in condensed matter physics for several decades. The presence of a chiral metallic edge mode that is robust to disorder and interactions at the boundary of 2D bulk Fermi liquid in strong magnetic field is a key feature of the quantum Hall effect. The existence of such an edge and the corresponding nontrivial topology of the bulk can be detected by computing a bulk topological number\cite{TKNN,Haldane,Bernevig2006,Bernevig2006s,Fu2007,Fu2007s,Bernevig2013,Ryu2008,Ryu2010,Moore2007,Alexandradinata,Fang2012,Fang2014,Lindner2011,Dzero2010}. Yet, in a more general sense, the robust gapless edge states within the quantum Hall effect are well-known to result in a U(1) chiral anomaly\cite{Adler1969,Bell,tHooft,Ishikawa1984,Laughlin1981}. The presence of anomalies in an edge theory and the resultant charge pumping imply the edge lives on the boundary of a higher dimensional manifold. In principle, the concept of quantum anomalies may be extended to characterize topological phases in the presence of interactions as the coefficients of the anomalies, which are quantized, are known to be stable against interactions\cite{Qi2013}.

Recent studies\cite{Ryu2012,Ryu2013,Hsieh2014,Cappelli2013,Shinsei2015} have proposed that an analysis of the anomalies in gapless $(1+1)$-D theories can also indicate the presence of a topological phase in $(2+1)$-D dimensions. This is based on the fact that if the edge theory has non-trivial response to certain transformations, which in $(1+1)$-D are chosen to be modular transformations, then the edge theory cannot be consistent on the $(1+1)$-D manifold and manifests itself as the edge of a $(2+1)$-D system. This method is known to give the correct results for chiral edge states in $(1+1)$-D, as well as for some more complex gapless edges involving spatial mirror symmetries \cite{Yao2013}. A necessary step is to extend this method beyond $(1+1)$-D to higher space dimension. In this letter, we extend concept of modular transformations of gapless free fermion theories beyond $(1+1)$-D to examine higher dimensional edge theories. We show that the complex free fermions in both $(2+1)$-D Dirac and $(3+1)$-D chiral edge theories are modularly invariant. However, when an external magnetic field is minimally coupled to the edge, the resultant Weyl modes show that a modular anomaly arises in the $(3+1)$-D edge theory indicating the presence of $(4+1)$-D quantum Hall effect. We further show using modular transformations that the edge theory of $(4+1)$-D insulators with the spacetime inversion symmetry$(P*T)$ and the fermion number parity symmetry for each spin becomes modular invariant when $8$ copies of the edges exist.

To begin, consider a relativistic conformal field theory (CFT) defined in a $(1+1)$-D compact space manifold $T^1\times T^1$ where $T^1$ is a torus (a circle in 1D). On such a space, the theory can exhibit invariance at a classical level under modular transformations\cite{Polchinski}. However, interesting cases arise when theories are not invariant under modular transformations resulting in the accumulation of an additional anomalous phase. The resultant anomaly is referred to as a large gravitational anomaly in the sense that it cannot be generated via continuous deformation of the original action\cite{Witten1985,Ryu2012}.  The modular group is defined as the group of linear fractional transformations of the upper half of the complex plane in which $\tau=L_0/L_1$ where $L_0$ and $L_1$ are the periods of the space and time coordinates respectively. $\tau$ transforms under the modular transformation:

%In particular, if the $(1+1)$-D compact space manifold has a chiral fermion living on it, the partition function of this chiral fermion cannot be made modular invariant, signaling the fact that the fermion lives at the edge of a higher-dimensional manifold.%

\bea
\label{Eq:Mod}
\tau'=\frac{a\tau+b}{c\tau+d},
\eea
where $a,b,c,d$ are integers satisfying $ad-bc=1$. The modular group is isomorphic to the projective special linear group $PSL(\mathbb{Z},2)$ \cite{Polchinski}. In $(1+1)$-D, the generators of the group are $S:\tau\rightarrow-1/\tau$ and $T:\tau\rightarrow \tau+1$. $S$ and $T$ act on the periods of each coordinate by $S:(L_0,L_1)\rightarrow(-L_1,L_0)$ and $T:(L_0,L_1)\rightarrow(L_0+L_1,L_1)$. To generalize modular transformation to higher dimensions, we consider the group generated by two generators, which they act on the periods of each coordinate as, $S:(L_0,L_1,L_2)\rightarrow(L_1,L_2,L_0)$, $T:(L_0,L_1,L_2)\rightarrow(L_0+L_1,L_1,L_2)$ in $(2+1)$-D, and $S:(L_0,L_1,L_2,L_3)\rightarrow(-L_1,L_2,L_3,L_0)$ and $T:(L_0,L_1,L_2,L_3)\rightarrow(L_0+L_1,L_1,L_2,L_3)$ in $(3+1)$-D. In this case, the generalized modular transformation is then isomorphic to $PSL(\mathbb{Z},d)$ (See Appendix. A). With these definitions, we consider the action of the modular group on the partition function in $(1+1)$-D\cite{Cappelli1997}, which is well-known to possess an anomaly. The most direct method to see the anomaly under the modular transformation is to calculate the partition function explicitly and apply the transformation. The partition function of $(1+1)$-D edge can be obtained in a well-regularized form as (for detailed calculation, see Appendix. B)
\begin{gather}
Z_{\lambda_0\lambda_1}=[e^{2\pi i(1/2-\lambda_0)(1/2-\lambda_1)}q^{(\lambda_1^2-\lambda_1+1/6)/2}]
\nonumber\\\times[(1-\omega)\prod_{n_1=1}^{\infty}(1-\omega q^{n_1})(1-\omega^{-1}q^{n_1})]
\end{gather}
where $\omega=q^{\lambda_1} e^{2\pi i \lambda_0}$, $q=e^{2\pi i \tau}$. $\lambda_0,\lambda_1=0(1/2)$ refers to the periodic (anti-periodic) boundary condition of the time and space coordinate directions respectively. By explicitly applying the modular transform, one derives the modular anomaly\cite{Ryu2013},
\begin{gather}
T[Z(\tau)_{\lambda_0\lambda_1}]=e^{i\pi(\lambda_1^2-\lambda_1+1/6)}Z(\tau)_{\lambda_0'\lambda_1'}
\\
\nonumber
S[Z(\tau)_{\lambda_0\lambda_1}]=e^{i2\pi(\lambda_1-1/2)(\lambda_0-1/2))}Z(\tau)_{\lambda_0'\lambda_1'}
\end{gather}
$\lambda'$ is the transformed boundary conditions under the modular transformation(See Appendix A). The sign of anomalous phase flips if the chirality of the $(1+1)$-D mode is reversed. Therefore, the combination of two edges of opposite chirality achieves modular invariance\cite{Cappelli2013}. This result is consistent with the fact that two opposite chiral edges can be gapped out by adding mass term. However, it is also possible to achieve modular invariance with finite copies of the same chirality\cite{Ryu2012}.

\begin{figure}
\includegraphics[width=0.4\textwidth]{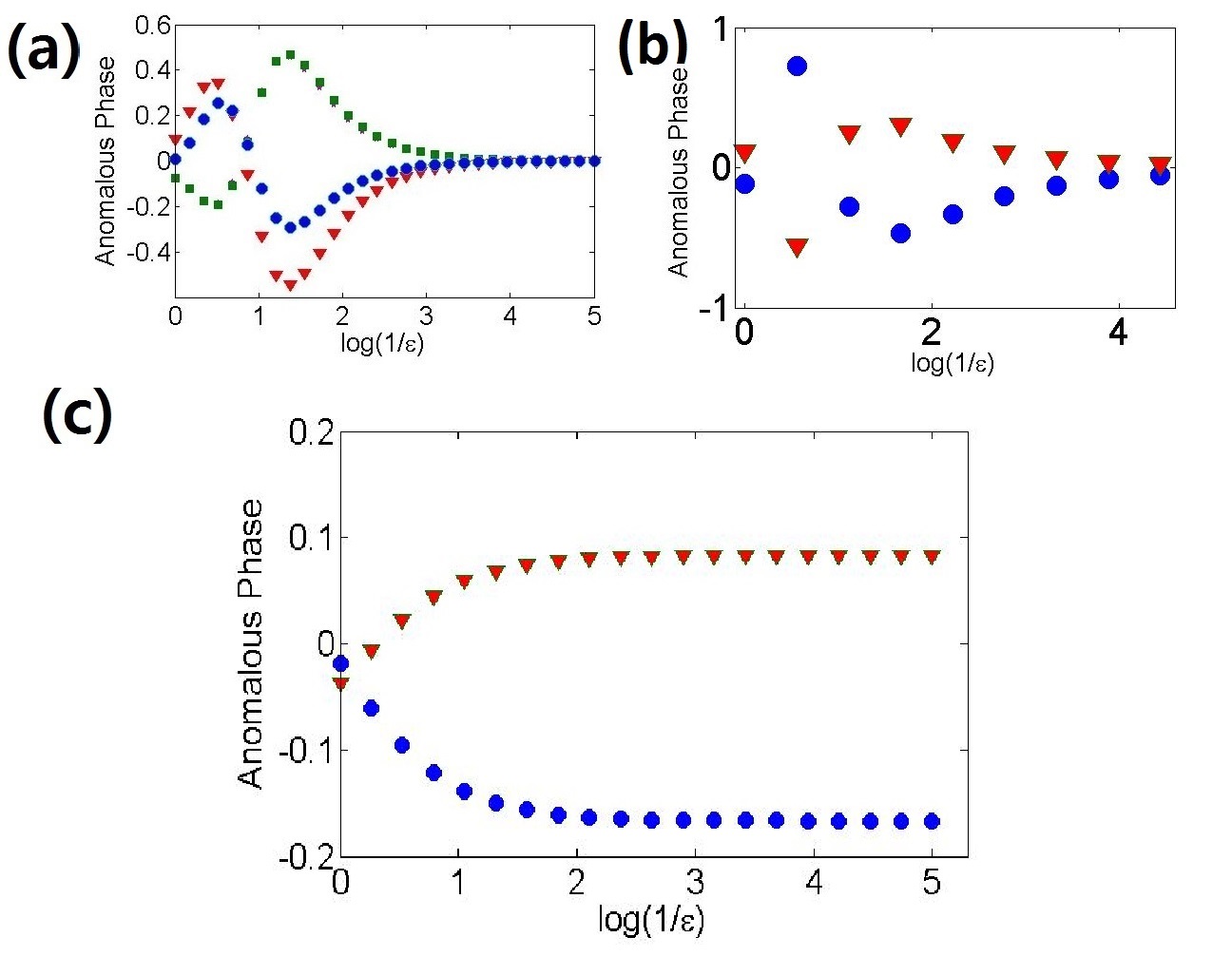}
\caption{\label{fig:FIG1} Calculation of numerical regularization scheme for: (a) $T$ transformation of (2+1)-D chiral edge. (b)(3+1)-D (c)(3+1)-D with magnetic field. Each lines represent different values of boundary condition. In (a), blue circles, red triangles and green squares represent $(\lambda_1,\lambda_2)=(0,0),(0.5,0),(0.5,0.5)$ respectively. In (b), blue circles and red triangles represent $(\lambda_1,\lambda_2,\lambda_3)=(0,0,0),(0.5,0,0)$. In (c), blue circles and red triangles represent $(\lambda_1,\lambda_2,\lambda_3)=(0.5,0,0),(0,0,0)$. We include sufficient numbers of high energy states within each calculation of the energy cutoff until the anomalous phase value converges. In (a) and (b) all choices of boundary conditions converge to zero indicating modular invariance. When the magnetic field is inserted, the anomaly approaches to the value(1/6,-1/12) in accordance with Eq. \eqref{Eq:Mod2} with $N_\phi=1$. Details of the numerical calculation method are provided in Appendix. D).}
\end{figure}
Now we wish to elucidate higher dimensional gapless edges, thus we examine $(2+1)$-D and $(3+1)$-D edge theories where the action is given by
\begin{gather}
\label{Eq:action}
\mathcal{S}=\int d^d x \bar\psi (\partial_\tau +\sigma\cdot k )\psi
\end{gather}
In contrast to $(1+1)$-D, we cannot simply perform the transformation of the partition function since an expression of the well-regularized partition function is not available. We can understand the failure of the regularization more clearly by applying the zeta function regularization method\cite{Hawking1977,Actor} to higher dimensional edge theories. In given edge theory, the expression of the unregularized partition function contains a summation of the energy eigenvalues over all states. In $(2+1)$-D, we have $\sum_{k_x,k_y} \sqrt{k_x^2+k_y^2}$ and in $(3+1)$-D, $\sum_{k_x,k_y,k_z} \sqrt{k_x^2+k_y^2+k_z^2}$ . When the sum is divergent, a successful zeta function regularization should utilize analytic continuation to assign a finite value to the divergent sum. Unfortunately, this is difficult since the Epstein-Hurwitz zeta (EZ) functions in $(3+1)$-D, $\zeta(\epsilon)=\sum_{n_1,n_2,n_3} ({n_1}^2+{n_2}^2+{n_3}^2)^{-\epsilon}$, and $(2+1)$-D, $\sum_{{n_1},{n_2}} ({n_1}^2+{n_2}^2)^{-\epsilon}$, are meromorphic at $\epsilon=-1/2$\cite{Elizalde}, which forbids assigning a finite value to the summation of energy eigenvalues. To circumvent this issue, we instead focus on the change of path integral measure\cite{Fujikawa1979,Fujikawa1980}. The calculation of the change of the measure only requires EZ function at $\epsilon=0$ and $\epsilon=-1$, which have well defined finite values (See Appendix C).

To calculate the change of measure, we work on the Fourier transformed field basis:
\begin{gather}
\psi_{\boldsymbol{\lambda}}(\vec{x},\vec{s})=\sum_{\vec{n},\vec{s}}a_{\vec{n},s} \Phi_{\vec{n}+\boldsymbol{\lambda}}(\vec{x})\chi_{\vec{s}}
\end{gather}
$\vec{n}=(n_0,n_1,n_2,n_3)$ are integers, which $n_i$ refers the frequency of $i$-th direction in the Fourier transformed basis. $\boldsymbol{\lambda}=(\lambda_0,\lambda_1,\lambda_2,\lambda_3)$ are the boundary conditions, which $\lambda_i=0(1/2)$ refers to the periodic (anti-periodic) boundary condition. We simplify the notations by defining $\widetilde{\vec{n}}=\vec{n}+\boldsymbol{\lambda}$. In other words,
\begin{gather}
\Phi_{(\widetilde{n}_0,\widetilde{n}_1,\widetilde{n}_2,\widetilde{n}_3)}(x)
=e^{2\pi i (\sum_{i=0}^4(n_i+\lambda_i)x_i)}
\end{gather}
and $\chi_\vec{s}$($\vec{s}=\pm$) is a two component spinor such that $\chi_{+}=(1,0)^T$,$\chi_{-}=(0,1)^T$. By following the transformation rule in appendix A, we represent the change of coefficient $a'$ under modular transformation as,
\begin{gather}
a'_{\vec{n}',\vec{s}'}
=\int d^d x \Phi^\dagger_{\vec{n}'+\boldsymbol{\lambda}'}\chi_{\vec{s}}^\dagger \psi_{\boldsymbol{\lambda}}(A^Tx)
\\
\nonumber
=\sum_{\vec{n},\vec{s}}[ \int d^d x \Phi^\dagger_{\vec{n}'+\boldsymbol{\lambda}'}(\vec{x}) \Phi_{\boldsymbol{\vec{n}+\lambda}}(A^T \vec{x}) ][\chi_{\vec{s}'}^\dagger \chi_{\vec{s}}]a_{\vec{n},\vec{s}},
\end{gather}
 where $A$ is the matrix representations of the generators. The above equation leads us to define the transformation matrix $C$ between $a_{\vec{n},\vec{s}}$ and $a'_{\vec{n},\vec{s}}$.
\begin{gather}
\label{defC}
C_{\vec{n}',\vec{n},\vec{s}',\vec{s}}=\int d^d x \Phi^\dagger_{\vec{n}'+\boldsymbol{\lambda}'}(\vec{x})\Phi_{\vec{n}+\boldsymbol{\lambda}}(A^T \vec{x})[\chi^\dagger_{\vec{s}'} \chi_{\vec{s}}]
\end{gather}
In terms of Fourier transformed field basis, the change of path integral measure is given by,
\begin{gather}
D\bar\psi' D\psi'= D\bar a' Da'= D\bar a Da ~det(C)^{-2}
\end{gather}
We treat  $\psi$ and $\bar\psi$ independently, hence we obtain an additional contribution of $-2$ sign from the Grassman algebra. In ($3+1$)-D, each momentum mode $\Phi$ transforms under modular transformation by (Appendix A),
\begin{gather}
\label{fourier}
T[\Phi_{\widetilde{n}_0,\widetilde{n}_1,\widetilde{n}_2,\widetilde{n}_3}]
=\Phi_{\widetilde{n}_0+\widetilde{n}_1,\widetilde{n}_1,\widetilde{n}_2,\widetilde{n}_3},
\\
\nonumber
S[\Phi_{\widetilde{n}_0,\widetilde{n}_1,\widetilde{n}_2,\widetilde{n}_3}]
=\Phi_{-\widetilde{n}_1,\widetilde{n}_2,\widetilde{n}_3,\widetilde{n}_0}.
\end{gather}
To calculate $Det(C)$, we select a basis that diagonalizes $C$. We define the basis as linear combinations of modes under successive applications of $T$ and $S$ as,
\begin{gather}
\label{newbasisT}
\ket{\theta,\overrightarrow{n}}_{\boldsymbol{\lambda}}\eta_{s,\overrightarrow{\widetilde{n}}}
=\Phi_{\boldsymbol{\lambda}} \sum_{n_0=0}^{n_1-1}\sum_{j=-\infty}^{\infty} e^{2\pi i (\widetilde{n}_0+{n_1}j)\theta} T^j[\Phi_{n_0,\overrightarrow{n}}]\eta_{s,\overrightarrow{\widetilde{n}}}
\\
\nonumber
\ket{\phi,\vec{n}}_{\boldsymbol{\lambda}}\chi_{s}=\Phi_{\boldsymbol{\lambda}-1/2}
\sum_{j=0}^{N-1} e^{2\pi i j\phi/N} S^j[\Phi_{\vec{n+1/2}}]\chi_{s},
\end{gather}
where $\overrightarrow{n}$ is the vector of the frequencies in spatial directions. $N$ is the order of $S$ such that $\Phi_{n_0,n_1,n_2,n_3}$ returns to the original mode  under $N$ application of $S$. In $(1+1),(2+1),(3+1)$-D, $N=4,3,8$ respectively, except for modes $\Phi_{n_0,n_0,n_0}$ in $(2+1)$-D. $\theta\in[0,1)$ and $\phi\in \lbrace -(\frac{N}{2}-1),-\frac{N-2}{2},..,\frac{N}{2} \rbrace$(In $(2+1)$-D,$\phi\in \lbrace -1,0,1 \rbrace$). To avoid double counting of the basis for $S$, we restrict the momentum indices to $n_0,n_1\geq 0$ in $(1+1)$-D, $n_0\geq n_1 \geq n_2$ in $(2+1)$-D, and $n_0,n_1,n_2 \geq 0$  in $(3+1)$-D. $\eta$ is the spinor which diagonalizes the Hamiltonian simultaneously. Then the basis satisfies
\begin{gather}
\label{eigenT}
T\ket{\theta,\overrightarrow{n}}_{\boldsymbol{\lambda}}=e^{-2\pi i \widetilde{n}_1\theta}\ket{\theta,\overrightarrow{n}}_{T[\boldsymbol{\lambda}]}
\\
\nonumber
S\ket{\phi,\vec{n}}_{\boldsymbol{\lambda}}=e^{-2\pi i \phi/N}\ket{\phi,\vec{n}}_{S[\boldsymbol{\lambda}]}
\end{gather}
In $(2+1)$-D, the $C$ matrix, using the new basis for $T$ and $S$, is a diagonal matrix given by
\begin{gather}
C^{2D}_{T,\{\theta,n_1,n_2,\theta',n_1',n_2'\}}=(e^{-2\pi i \widetilde{n}_1\theta})\delta(\theta-\theta')\delta_{\overrightarrow{n},\overrightarrow{n}'}
\nonumber
\\
C^{2D}_{S,\{\phi,n_0,n_1,n_2,\phi',n_0',n_1',n_2'\}}=(e^{-2\pi i \phi/N})\delta_{\phi,\phi'}\delta_{\vec{n},\vec{n}'}.
\end{gather}
To regulate the determinant of $C$, we use the Epstein-Hurwitz type zeta function regulator that has the same form as energy dispersion, $|p|^{-\epsilon}$, where $\epsilon$, in this instance, is a scale which cuts off the high energy states. In $(2+1)$-D, we find the phase of the regulated determinants of $C_S$ and the submatrix of $C_T$ with positive momenta ($n_1,n_2>0$) to be (See Appendix C):
\begin{gather}
\label{c2d}
arg(det(C^{2D}_{S}))=0
\\
\nonumber
arg(det(C^{2D}_{T,\{n_1,n_2>0\}}))=-\frac{\pi L_x^2}{2}\partial_{\lambda_1}E_2(-1,\lambda_1,\lambda_2).
\end{gather}
In Eq. (\ref{c2d}),  $E_2(\epsilon,\lambda_1,\lambda_2)=\sum_{n_1,n_2>0}(\widetilde{n}_1^2+\widetilde{n}_2^2)^{-\epsilon}$ is the $2$D EZ function of positive integer plane. We can expand the sum to full integer momentum sectors by using following identity of EZ(Appendix C),
\begin{gather}
\label{ezid}
\sum_{n_1,n_2=-\infty}^{\infty}(\widetilde{n}_1^2+\widetilde{n}_2^2)^{-\epsilon}=\sum_{c_{1,2}=\lambda_{1,2},1-\lambda_{1,2}}E_2(s,c_1,c_2)
\\
\nonumber
E_2(\epsilon,\lambda_1,\lambda_2)=-E_2(\epsilon,1-\lambda_1,\lambda_2)=-E_2(\epsilon,\lambda_1,1-\lambda_2)
\end{gather}
From Eq. (\ref{c2d}) and Eq. (\ref{ezid}), $arg(det(C_T^{2D}))$ of negative momentum sector exactly cancels the contribution of positive momentum sector resulting in modular invariance of the $(2+1)$-D edge theory. We confirm our result numerically in Fig. \ref{fig:FIG1}(a) via a calculation of the Casmir energy (see Appendix D).The cancellation of the modular anomaly is not surprising as one may represent a gapless theory in $(2+1)$-D on a lattice indicating that a higher dimensional bulk is not required to regularize a $(2+1)$-D theory. Further, this indicates that $(2+1)$-D gapless theory can be generically gapped out without time-reversal symmetry. However, by adding symmetry constraints, a modular anomaly can be found\cite{Shinsei2015}.

In contrast to $(2+1)$-D, the Nielsen-Ninomiya (NN) theorem in $(3+1)$-D suggests that the chiral edge of even dimension cannot be written without the aid of bulk theory\cite{Nielsen1981}. Therefore, it is natural to expect an anomalous contribution in even dimensions even without symmetry projection. In a similar manner with $(2+1)$-D, the modular transformations, $S$ and $T$, of the transformation matrix, $C$, in $(3+1)$-D are given by(See Appendix C),
\begin{gather}
C^{3D}_{T,\{\theta,n_1,n_2,n_3,\theta',n_1',n_2',n_3'\}}=(e^{-2\pi i \widetilde{n}_1\theta})\delta(\theta-\theta')\delta_{\overrightarrow{n},\overrightarrow{n}'}.
\\
\nonumber
C^{3D}_{S,\{\phi,n_0,n_1,n_2,n_3,\phi',n_0',n_1',n_2',n_3'\}}=(e^{-2\pi i \phi/N})\delta_{\phi,\phi'}\delta_{\vec{n},\vec{n}'}.
\end{gather}
The determinants of $C_S$ and a submatrix of $C_T$ with positive momenta ($n_1,n_2,n_3>0$) are given by,
\begin{gather}
arg(det(C^{3D}_{S}))=-2\pi(1/2-\lambda_0) \sum_{c_3=\lambda_3,1-\lambda_3}E_3(0,\lambda_1,\lambda_2,c_3)
\nonumber
\\
arg(det(C^{3D}_{T,\{n_1,n_2,n_3>0\}}))=-\frac{\pi L_x^2}{2}\partial_{\lambda_1}E_3(-1,\lambda_1,\lambda_2,\lambda_3)
\end{gather}
where $E_3(\epsilon,\lambda_1,\lambda_2,\lambda_3)=\sum_{n_1,n_2,n_3>0}(\widetilde{n}_1^2+\widetilde{n}_2^2+\widetilde{n}_3^2)^{-\epsilon}$ is the $3$D EZ function on the positive integer plane. By expanding the sum to the full integer plane(See Appendix C), we again find the anomalous phase contribution cancels out and that we find that free gapless fermions in $(3+1)$-D are modular invariant. To confirm our result, we numerically calculate the Casmir energy of our $(3+1)$-D edge theory(See Fig. \ref{fig:FIG1}(b)) to again find the modular invariance of $(3+1)$-D Weyl fermions in agreement with the result obtained via the change in transformation matrix.

While chiral free fermions in $(3+1)$-D are modular anomaly free, attaching a background $U(1)$ gauge field changes the situation. Consider the chiral edge under a uniform magnetic field pointing out-of-plane in the $z$-direction thereby breaking the periodicity of the in-plane $x$ and $y$ coordinates. Therefore, the full modular transformation that is isomorphic to $PSL(\mathbb{Z},4)$ is no longer a good symmetry of the action, Eq. (\ref{Eq:action}). However, we can still safely consider $PSL(\mathbb{Z},2)$ acting on both $z$ and the time component as a subgroup of the original $PSL(\mathbb{Z},4)$. We write the Hamiltonian for this situation as,
\begin{equation}
\label{Eq:hamll}
H=(\overrightarrow{k}-\overrightarrow{A})\cdot \overrightarrow{\sigma},
\end{equation}
with magnetic vector potential $A$ written in the Landau gauge, $A=(0,-Bx,0)$. This Hamiltonian has two types of solutions. $E_{W(D)}$ is gapless(gapped) Landau level(LL).
\begin{equation}
\label{Eq:dispers}
E_{W}(k_3)=k_z,E_{D}(n,k_3)=\pm\sqrt{Bn+k_3^2},
\end{equation}
where $k_3=2\pi (n_3+\lambda_3)/L_z$ and $n$ is an positive integer. We can write the unregularized partition function to be
\begin{gather}
Z_{\lambda_0,\lambda_3}=[\prod_{k_3} (1- e^{2\pi i \lambda_0-\beta E_{W}(k_z)})
\\\nonumber\times\prod_{n,k_z}(1- e^{2\pi i \lambda_0+\beta E_D(n,k_z)})(1- e^{2\pi i \lambda_0-\beta E_D(n,k_z)})]^{N_\phi},
\end{gather}
where $N_\phi$ is the level degeneracy and $\omega=q^{\lambda_3} e^{2\pi i \lambda_0}$. After regularization, we find that the chiral modes contribute to the anomaly while gapped landau levels do not contribute as they are massive (See Appendix E). This reflects the fact that the regularized Casimir energy has no contribution from gapped states. Therefore, the modular transforms of the partition function of a $(3+1)$-D edge theory coupled to a $U(1)$ gauge field are
\begin{eqnarray}
\label{Eq:Mod2}
T[Z_{\boldsymbol{\lambda}}]=e^{iN_\phi\pi(\lambda_3^2-\lambda_3+1/6)}Z_{\boldsymbol{\lambda}'}
\\
\nonumber
S[Z_{\boldsymbol{\lambda}}]=e^{iN_\phi 2\pi(\lambda_3-1/2)(\lambda_0-1/2))}Z_{\boldsymbol{\lambda}'}
\end{eqnarray}
and clearly contain a modular anomaly that is proportional to $N_\phi$, which counts the number of $(1+1)$-D chiral modes. To confirm, we again look at the numerical calculation of the Casmir energy in Fig. \ref{fig:FIG1}(c) where we find that the anomalous phase value under $T$ transformation reproduces the transformation rules given in Eq. \eqref{Eq:Mod2}. Thus, the $(3+1)$-D chiral edge, when coupled to a background gauge field, contains a modular anomaly. In contrast to $(1+1)$-D, $(3+1)$-D chiral edge has charge pumping but only in conjunction with the magnetic field, in analogy to the chiral anomaly\cite{Nielsen1983}. Therefore, we conclude the presence of modular anomaly when $N_\phi \neq 0$, is a direct manifestation of quantum Hall effect of ($4+1$)-D.

Using the $(3+1)$-D chiral edge result enables us to extend our analysis to ($4+1$)-D insulators with the following two symmetries: the fermion parity of each spin component is preserved and the system is invariant under $(x,y,z,w,t)\rightarrow(-x,-y,-z,w,-t)$. Consider an open ($3+1$)-D surface with $w=const$, the second symmetry is equivalent to $P*T$, where $P$ is parity in 3D. The first symmetry, in the non-interacting case, ensures the decoupling between spin-up and spin-down sectors, so each sector is itself a $(4+1)$-D quantum Hall state with $N_\uparrow$ ($N_{\downarrow}$) Weyl nodes on the surface. $P*T$ on the surface ensures that $N_\uparrow=N_\downarrow\equiv{N_{edge}}$, as it maps (i) spin-up to spin-down and (ii) a positive monopole to a negative monopole. The topological classification without interaction is hence $\mathbb{Z}$. Since the fermion parity of each spin is separately conserved, the total partition function is\cite{Ryu2012,Cappelli2013}(See Appendix F),
\bea
Z_{total}=|\sum_{\lambda_0,\lambda_3=0,1/2} Z_{\lambda_0,\lambda_3}^{N_{edge}}|^2.
\eea
It is found that $Z_{total}$ is modular invariant under both $S$ and $T$ only when $N_{edge}=8/gcd(N_\phi,8)$. As $N_\phi$ can be any integer, the complete cancellation of the modular anomaly occurs when $N_{edge}=8$.

%Beyond the detection of the modular anomaly $(3+1)$-D edge with magnetic field, it is important to understand the classification. Here we study the modular invariance with fermion number parity symmetry($\mathbb{Z}_2$). The fermion number parity is especially of interest as it pertains to the case of topological superconductors\cite{Kitaev}, where the fermion number is only conserved $mod(2)$. We start by considering $N$ copies of the complex fermions along the edge coupled to the magnetic field that has $N_\phi$ Landau degeneracy. We enforce the symmetry by dividing Hilbert space into sectors labeled by fermion number parity and inserting the symmetry projection operator in the calculation of the partition function as
%\bea
%Z=tr_{\lambda}(Pq^{H_{\vec{\lambda}}})
%\eea
%where the projection operator is given by,
%\bea
%P_{chiral,\mathbb{Z}_2}=(1+(-1)^{N})/2
%\eea
%This is identical to considering summation of all possible cases of partition function with periodic and anti-periodic boundary conditions\cite{Ryu2012}. We find that the chiral theory can achieve modular covariance when $N=8/gcd(N_\phi,8)$ and modular invariance is achieved when $N=24/gcd(N_\phi,24)$ (For detailed calculation, see Appendix F). Therefore, when $N=8(24)$, modular covariance (invariance) is achieved regardless of $N_\phi$

In conclusion, we have generalized modular transformation in $(1+1)$-D CFT to higher dimensional edge theory with use of $PSL(\mathbb{Z},d)$ group supported by numerical calculations of the Casmir energies. We have shown the gapless free fermion theories in $(2+1)$-D and $(3+1)$-D are modular invariant. We find a modular anomaly in $(3+1)$-D when the edge theory is coupled to a $U(1)$ gauge field resulting in a $(4+1)$-D quantum Hall effect. Moreover, we find that the edge theory of $(4+1)$-D insulator with spacetime inversion symmetry$(P*T)$ and fermion number parity symmetry for each spin achieves modular invariant when $N_{edge}=8$.

\begin{acknowledgements}
MJG would like to thank Shinsei Ryu for enlightening conversations. MJP would like to thank Thomas Faulkner for helpful discussions.
 MJP, CF, and MJG were supported by ONR - N0014-11-1-0123. MJG and MJP were supported by NSF-CAREER EECS-1351871. BAB was supported by NSF CAREER DMR-095242, ONR - N00014\text{-}11\text{-}1-0635, MURI\text{-}130\text{-}6082,  NSF-MRSEC DMR-0819860, DARPA under SPAWAR Grant No.: N66001\text{-}11\text{-}1-4110, Packard Foundation and Keck grant.
\end{acknowledgements}

\bibliography{modularbib}

\begin{thebibliography}{39}%
\makeatletter
\providecommand \@ifxundefined [1]{%
 \@ifx{#1\undefined}
}%
\providecommand \@ifnum [1]{%
 \ifnum #1\expandafter \@firstoftwo
 \else \expandafter \@secondoftwo
 \fi
}%
\providecommand \@ifx [1]{%
 \ifx #1\expandafter \@firstoftwo
 \else \expandafter \@secondoftwo
 \fi
}%
\providecommand \natexlab [1]{#1}%
\providecommand \enquote  [1]{``#1''}%
\providecommand \bibnamefont  [1]{#1}%
\providecommand \bibfnamefont [1]{#1}%
\providecommand \citenamefont [1]{#1}%
\providecommand \href@noop [0]{\@secondoftwo}%
\providecommand \href [0]{\begingroup \@sanitize@url \@href}%
\providecommand \@href[1]{\@@startlink{#1}\@@href}%
\providecommand \@@href[1]{\endgroup#1\@@endlink}%
\providecommand \@sanitize@url [0]{\catcode `\\12\catcode `\$12\catcode
  `\&12\catcode `\#12\catcode `\^12\catcode `\_12\catcode `\%12\relax}%
\providecommand \@@startlink[1]{}%
\providecommand \@@endlink[0]{}%
\providecommand \url  [0]{\begingroup\@sanitize@url \@url }%
\providecommand \@url [1]{\endgroup\@href {#1}{\urlprefix }}%
\providecommand \urlprefix  [0]{URL }%
\providecommand \Eprint [0]{\href }%
\providecommand \doibase [0]{http://dx.doi.org/}%
\providecommand \selectlanguage [0]{\@gobble}%
\providecommand \bibinfo  [0]{\@secondoftwo}%
\providecommand \bibfield  [0]{\@secondoftwo}%
\providecommand \translation [1]{[#1]}%
\providecommand \BibitemOpen [0]{}%
\providecommand \bibitemStop [0]{}%
\providecommand \bibitemNoStop [0]{.\EOS\space}%
\providecommand \EOS [0]{\spacefactor3000\relax}%
\providecommand \BibitemShut  [1]{\csname bibitem#1\endcsname}%
\let\auto@bib@innerbib\@empty
%</preamble>
\bibitem [{\citenamefont {v.~Klitzing}\ \emph {et~al.}(1980)\citenamefont
  {v.~Klitzing}, \citenamefont {Dorda},\ and\ \citenamefont
  {Pepper}}]{klitzing1980}%
  \BibitemOpen
  \bibfield  {author} {\bibinfo {author} {\bibfnamefont {K.}~\bibnamefont
  {v.~Klitzing}}, \bibinfo {author} {\bibfnamefont {G.}~\bibnamefont {Dorda}},
  \ and\ \bibinfo {author} {\bibfnamefont {M.}~\bibnamefont {Pepper}},\
  }\href@noop {} {\bibfield  {journal} {\bibinfo  {journal} {Phys. Rev. Lett}\
  }\textbf {\bibinfo {volume} {45}},\ \bibinfo {pages} {494} (\bibinfo {year}
  {1980})}\BibitemShut {NoStop}%
\bibitem [{\citenamefont {Thouless}\ \emph {et~al.}(1982)\citenamefont
  {Thouless}, \citenamefont {Kohmoto}, \citenamefont {Nightingale},\ and\
  \citenamefont {den Nijs}}]{TKNN}%
  \BibitemOpen
  \bibfield  {author} {\bibinfo {author} {\bibfnamefont {D.~J.}\ \bibnamefont
  {Thouless}}, \bibinfo {author} {\bibfnamefont {M.}~\bibnamefont {Kohmoto}},
  \bibinfo {author} {\bibfnamefont {M.~P.}\ \bibnamefont {Nightingale}}, \ and\
  \bibinfo {author} {\bibfnamefont {M.}~\bibnamefont {den Nijs}},\ }\href
  {\doibase 10.1103/PhysRevLett.49.405} {\bibfield  {journal} {\bibinfo
  {journal} {Phys. Rev. Lett.}\ }\textbf {\bibinfo {volume} {49}},\ \bibinfo
  {pages} {405} (\bibinfo {year} {1982})}\BibitemShut {NoStop}%
\bibitem [{\citenamefont {Haldane}(1988)}]{Haldane}%
  \BibitemOpen
  \bibfield  {author} {\bibinfo {author} {\bibfnamefont {F.~D.~M.}\
  \bibnamefont {Haldane}},\ }\href {\doibase 10.1103/PhysRevLett.61.2015}
  {\bibfield  {journal} {\bibinfo  {journal} {Phys. Rev. Lett.}\ }\textbf
  {\bibinfo {volume} {61}},\ \bibinfo {pages} {2015} (\bibinfo {year}
  {1988})}\BibitemShut {NoStop}%
\bibitem [{\citenamefont {Bernevig}\ and\ \citenamefont
  {Zhang}(2006)}]{Bernevig2006}%
  \BibitemOpen
  \bibfield  {author} {\bibinfo {author} {\bibfnamefont {B.~A.}\ \bibnamefont
  {Bernevig}}\ and\ \bibinfo {author} {\bibfnamefont {S.-C.}\ \bibnamefont
  {Zhang}},\ }\href@noop {} {\bibfield  {journal} {\bibinfo  {journal} {Phys.
  Rev. Lett.}\ }\textbf {\bibinfo {volume} {96}},\ \bibinfo {pages} {106802}
  (\bibinfo {year} {2006})}\BibitemShut {NoStop}%
\bibitem [{\citenamefont {B.~A.~Bernevig}\ and\ \citenamefont
  {Zhang}(2006)}]{Bernevig2006s}%
  \BibitemOpen
  \bibfield  {author} {\bibinfo {author} {\bibfnamefont {T.~H.}\ \bibnamefont
  {B.~A.~Bernevig}}\ and\ \bibinfo {author} {\bibfnamefont {S.-C.}\
  \bibnamefont {Zhang}},\ }\href@noop {} {\bibfield  {journal} {\bibinfo
  {journal} {Science}\ }\textbf {\bibinfo {volume} {314}},\ \bibinfo {pages}
  {1757} (\bibinfo {year} {2006})}\BibitemShut {NoStop}%
\bibitem [{\citenamefont {Fu}\ \emph {et~al.}(2007)\citenamefont {Fu},
  \citenamefont {Kane},\ and\ \citenamefont {Mele}}]{Fu2007}%
  \BibitemOpen
  \bibfield  {author} {\bibinfo {author} {\bibfnamefont {L.}~\bibnamefont
  {Fu}}, \bibinfo {author} {\bibfnamefont {C.~L.}\ \bibnamefont {Kane}}, \ and\
  \bibinfo {author} {\bibfnamefont {E.~J.}\ \bibnamefont {Mele}},\ }\href@noop
  {} {\bibfield  {journal} {\bibinfo  {journal} {Phys. Rev. Lett.}\ }\textbf
  {\bibinfo {volume} {98}},\ \bibinfo {pages} {106801} (\bibinfo {year}
  {2007})}\BibitemShut {NoStop}%
\bibitem [{\citenamefont {Fu}\ and\ \citenamefont {Kane}(2007)}]{Fu2007s}%
  \BibitemOpen
  \bibfield  {author} {\bibinfo {author} {\bibfnamefont {L.}~\bibnamefont
  {Fu}}\ and\ \bibinfo {author} {\bibfnamefont {C.~L.}\ \bibnamefont {Kane}},\
  }\href@noop {} {\bibfield  {journal} {\bibinfo  {journal} {Phys. Rev. B}\
  }\textbf {\bibinfo {volume} {76}},\ \bibinfo {pages} {045302} (\bibinfo
  {year} {2007})}\BibitemShut {NoStop}%
\bibitem [{\citenamefont {Hughes}\ and\ \citenamefont
  {Bernevig}(2013)}]{Bernevig2013}%
  \BibitemOpen
  \bibfield  {author} {\bibinfo {author} {\bibfnamefont {T.~L.}\ \bibnamefont
  {Hughes}}\ and\ \bibinfo {author} {\bibfnamefont {B.~A.}\ \bibnamefont
  {Bernevig}},\ }\href@noop {} {\emph {\bibinfo {title} {Topological Insulators
  and Topological Superconductors}}}\ (\bibinfo  {publisher} {Princeton
  University Press},\ \bibinfo {year} {2013})\BibitemShut {NoStop}%
\bibitem [{\citenamefont {Schnyder}\ \emph {et~al.}(2008)\citenamefont
  {Schnyder}, \citenamefont {Ryu}, \citenamefont {Furusaki},\ and\
  \citenamefont {Ludwig}}]{Ryu2008}%
  \BibitemOpen
  \bibfield  {author} {\bibinfo {author} {\bibfnamefont {A.~P.}\ \bibnamefont
  {Schnyder}}, \bibinfo {author} {\bibfnamefont {S.}~\bibnamefont {Ryu}},
  \bibinfo {author} {\bibfnamefont {A.}~\bibnamefont {Furusaki}}, \ and\
  \bibinfo {author} {\bibfnamefont {A.~W.~W.}\ \bibnamefont {Ludwig}},\
  }\href@noop {} {\bibfield  {journal} {\bibinfo  {journal} {Phys. Rev. B}\
  }\textbf {\bibinfo {volume} {78}},\ \bibinfo {pages} {195125} (\bibinfo
  {year} {2008})}\BibitemShut {NoStop}%
\bibitem [{\citenamefont {Ryu}\ \emph {et~al.}(2010)\citenamefont {Ryu},
  \citenamefont {A.~Schnyder},\ and\ \citenamefont {Ludwig}}]{Ryu2010}%
  \BibitemOpen
  \bibfield  {author} {\bibinfo {author} {\bibfnamefont {S.}~\bibnamefont
  {Ryu}}, \bibinfo {author} {\bibfnamefont {A.~F.}\ \bibnamefont
  {A.~Schnyder}}, \ and\ \bibinfo {author} {\bibfnamefont {A.~W.~W.}\
  \bibnamefont {Ludwig}},\ }\href@noop {} {\bibfield  {journal} {\bibinfo
  {journal} {New J. Phys.}\ }\textbf {\bibinfo {volume} {12}},\ \bibinfo
  {pages} {065010} (\bibinfo {year} {2010})}\BibitemShut {NoStop}%
\bibitem [{\citenamefont {Moore}\ and\ \citenamefont
  {Balents}(2007)}]{Moore2007}%
  \BibitemOpen
  \bibfield  {author} {\bibinfo {author} {\bibfnamefont {J.~E.}\ \bibnamefont
  {Moore}}\ and\ \bibinfo {author} {\bibfnamefont {L.}~\bibnamefont
  {Balents}},\ }\href@noop {} {\bibfield  {journal} {\bibinfo  {journal} {Phys.
  Rev. B}\ }\textbf {\bibinfo {volume} {75}},\ \bibinfo {pages} {121306(R)}
  (\bibinfo {year} {2007})}\BibitemShut {NoStop}%
\bibitem [{\citenamefont {Alexandradinata}\ \emph {et~al.}(2014)\citenamefont
  {Alexandradinata}, \citenamefont {Fang}, \citenamefont {Gilbert},\ and\
  \citenamefont {Bernevig}}]{Alexandradinata}%
  \BibitemOpen
  \bibfield  {author} {\bibinfo {author} {\bibfnamefont {A.}~\bibnamefont
  {Alexandradinata}}, \bibinfo {author} {\bibfnamefont {C.}~\bibnamefont
  {Fang}}, \bibinfo {author} {\bibfnamefont {M.~J.}\ \bibnamefont {Gilbert}}, \
  and\ \bibinfo {author} {\bibfnamefont {B.~A.}\ \bibnamefont {Bernevig}},\
  }\href {\doibase 10.1103/PhysRevLett.113.116403} {\bibfield  {journal}
  {\bibinfo  {journal} {Phys. Rev. Lett.}\ }\textbf {\bibinfo {volume} {113}},\
  \bibinfo {pages} {116403} (\bibinfo {year} {2014})}\BibitemShut {NoStop}%
\bibitem [{\citenamefont {Fang}\ \emph {et~al.}(2012)\citenamefont {Fang},
  \citenamefont {Gilbert},\ and\ \citenamefont {Bernevig}}]{Fang2012}%
  \BibitemOpen
  \bibfield  {author} {\bibinfo {author} {\bibfnamefont {C.}~\bibnamefont
  {Fang}}, \bibinfo {author} {\bibfnamefont {M.~J.}\ \bibnamefont {Gilbert}}, \
  and\ \bibinfo {author} {\bibfnamefont {B.~A.}\ \bibnamefont {Bernevig}},\
  }\href {\doibase 10.1103/PhysRevB.86.115112} {\bibfield  {journal} {\bibinfo
  {journal} {Phys. Rev. B}\ }\textbf {\bibinfo {volume} {86}},\ \bibinfo
  {pages} {115112} (\bibinfo {year} {2012})}\BibitemShut {NoStop}%
\bibitem [{\citenamefont {Fang}\ \emph {et~al.}(2014)\citenamefont {Fang},
  \citenamefont {Gilbert},\ and\ \citenamefont {Bernevig}}]{Fang2014}%
  \BibitemOpen
  \bibfield  {author} {\bibinfo {author} {\bibfnamefont {C.}~\bibnamefont
  {Fang}}, \bibinfo {author} {\bibfnamefont {M.~J.}\ \bibnamefont {Gilbert}}, \
  and\ \bibinfo {author} {\bibfnamefont {B.~A.}\ \bibnamefont {Bernevig}},\
  }\href {\doibase 10.1103/PhysRevLett.112.106401} {\bibfield  {journal}
  {\bibinfo  {journal} {Phys. Rev. Lett.}\ }\textbf {\bibinfo {volume} {112}},\
  \bibinfo {pages} {106401} (\bibinfo {year} {2014})}\BibitemShut {NoStop}%
\bibitem [{\citenamefont {Lindner}\ \emph {et~al.}(2011)\citenamefont
  {Lindner}, \citenamefont {Refael},\ and\ \citenamefont
  {Galitski}}]{Lindner2011}%
  \BibitemOpen
  \bibfield  {author} {\bibinfo {author} {\bibfnamefont {N.~H.}\ \bibnamefont
  {Lindner}}, \bibinfo {author} {\bibfnamefont {G.}~\bibnamefont {Refael}}, \
  and\ \bibinfo {author} {\bibfnamefont {V.}~\bibnamefont {Galitski}},\
  }\href@noop {} {\bibfield  {journal} {\bibinfo  {journal} {Nature Physics}\
  }\textbf {\bibinfo {volume} {7}},\ \bibinfo {pages} {490} (\bibinfo {year}
  {2011})}\BibitemShut {NoStop}%
\bibitem [{\citenamefont {Dzero}\ \emph {et~al.}(2010)\citenamefont {Dzero},
  \citenamefont {Sun}, \citenamefont {Galitski},\ and\ \citenamefont
  {Coleman}}]{Dzero2010}%
  \BibitemOpen
  \bibfield  {author} {\bibinfo {author} {\bibfnamefont {M.}~\bibnamefont
  {Dzero}}, \bibinfo {author} {\bibfnamefont {K.}~\bibnamefont {Sun}}, \bibinfo
  {author} {\bibfnamefont {V.}~\bibnamefont {Galitski}}, \ and\ \bibinfo
  {author} {\bibfnamefont {P.}~\bibnamefont {Coleman}},\ }\href@noop {}
  {\bibfield  {journal} {\bibinfo  {journal} {Phys. Rev. Lett.}\ }\textbf
  {\bibinfo {volume} {104}},\ \bibinfo {pages} {106408} (\bibinfo {year}
  {2010})}\BibitemShut {NoStop}%
\bibitem [{\citenamefont {Adler}(1969)}]{Adler1969}%
  \BibitemOpen
  \bibfield  {author} {\bibinfo {author} {\bibfnamefont {S.}~\bibnamefont
  {Adler}},\ }\href@noop {} {\bibfield  {journal} {\bibinfo  {journal} {Phys.
  Rev.}\ }\textbf {\bibinfo {volume} {5}},\ \bibinfo {pages} {177} (\bibinfo
  {year} {1969})}\BibitemShut {NoStop}%
\bibitem [{\citenamefont {Bell}\ and\ \citenamefont {Jackiw}(1969)}]{Bell}%
  \BibitemOpen
  \bibfield  {author} {\bibinfo {author} {\bibfnamefont {J.}~\bibnamefont
  {Bell}}\ and\ \bibinfo {author} {\bibfnamefont {R.}~\bibnamefont {Jackiw}},\
  }\href@noop {} {\bibfield  {journal} {\bibinfo  {journal} {Il Nuovo Cimento}\
  }\textbf {\bibinfo {volume} {60}},\ \bibinfo {pages} {47} (\bibinfo {year}
  {1969})}\BibitemShut {NoStop}%
\bibitem [{\citenamefont {'t~Hooft}(1976)}]{tHooft}%
  \BibitemOpen
  \bibfield  {author} {\bibinfo {author} {\bibfnamefont {G.}~\bibnamefont
  {'t~Hooft}},\ }\href@noop {} {\bibfield  {journal} {\bibinfo  {journal}
  {Phys. Rev. Lett.}\ }\textbf {\bibinfo {volume} {8}},\ \bibinfo {pages} {37}
  (\bibinfo {year} {1976})}\BibitemShut {NoStop}%
\bibitem [{\citenamefont {Ishikawa}(1984)}]{Ishikawa1984}%
  \BibitemOpen
  \bibfield  {author} {\bibinfo {author} {\bibfnamefont {K.}~\bibnamefont
  {Ishikawa}},\ }\href@noop {} {\bibfield  {journal} {\bibinfo  {journal}
  {Phys. Rev. Lett}\ }\textbf {\bibinfo {volume} {53}},\ \bibinfo {pages}
  {1615} (\bibinfo {year} {1984})}\BibitemShut {NoStop}%
\bibitem [{\citenamefont {Laughlin}(1981)}]{Laughlin1981}%
  \BibitemOpen
  \bibfield  {author} {\bibinfo {author} {\bibfnamefont {R.~B.}\ \bibnamefont
  {Laughlin}},\ }\href@noop {} {\bibfield  {journal} {\bibinfo  {journal}
  {Phys. Rev. B}\ }\textbf {\bibinfo {volume} {23}},\ \bibinfo {pages} {5632}
  (\bibinfo {year} {1981})}\BibitemShut {NoStop}%
\bibitem [{\citenamefont {Qi}\ \emph {et~al.}(2013)\citenamefont {Qi},
  \citenamefont {Witten},\ and\ \citenamefont {Zhang}}]{Qi2013}%
  \BibitemOpen
  \bibfield  {author} {\bibinfo {author} {\bibfnamefont {X.-L.}\ \bibnamefont
  {Qi}}, \bibinfo {author} {\bibfnamefont {E.}~\bibnamefont {Witten}}, \ and\
  \bibinfo {author} {\bibfnamefont {S.-C.}\ \bibnamefont {Zhang}},\ }\href@noop
  {} {\bibfield  {journal} {\bibinfo  {journal} {Phys. Rev. B}\ }\textbf
  {\bibinfo {volume} {87}},\ \bibinfo {pages} {134519} (\bibinfo {year}
  {2013})}\BibitemShut {NoStop}%
\bibitem [{\citenamefont {Ryu}\ and\ \citenamefont {Zhang}(2012)}]{Ryu2012}%
  \BibitemOpen
  \bibfield  {author} {\bibinfo {author} {\bibfnamefont {S.}~\bibnamefont
  {Ryu}}\ and\ \bibinfo {author} {\bibfnamefont {S.~C.}\ \bibnamefont
  {Zhang}},\ }\href@noop {} {\bibfield  {journal} {\bibinfo  {journal} {Phys.
  Rev. B}\ }\textbf {\bibinfo {volume} {85}},\ \bibinfo {pages} {245132}
  (\bibinfo {year} {2012})}\BibitemShut {NoStop}%
\bibitem [{\citenamefont {Sule}\ \emph {et~al.}(2013)\citenamefont {Sule},
  \citenamefont {Chen},\ and\ \citenamefont {Ryu}}]{Ryu2013}%
  \BibitemOpen
  \bibfield  {author} {\bibinfo {author} {\bibfnamefont {O.}~\bibnamefont
  {Sule}}, \bibinfo {author} {\bibfnamefont {X.}~\bibnamefont {Chen}}, \ and\
  \bibinfo {author} {\bibfnamefont {S.}~\bibnamefont {Ryu}},\ }\href@noop {}
  {\bibfield  {journal} {\bibinfo  {journal} {Phys. Rev. B}\ }\textbf {\bibinfo
  {volume} {88}},\ \bibinfo {pages} {075125} (\bibinfo {year}
  {2013})}\BibitemShut {NoStop}%
\bibitem [{\citenamefont {Hsieh}\ \emph {et~al.}(2014)\citenamefont {Hsieh},
  \citenamefont {Sule}, \citenamefont {Cho}, \citenamefont {Ryu},\ and\
  \citenamefont {Leigh}}]{Hsieh2014}%
  \BibitemOpen
  \bibfield  {author} {\bibinfo {author} {\bibfnamefont {C.}~\bibnamefont
  {Hsieh}}, \bibinfo {author} {\bibfnamefont {O.~M.}\ \bibnamefont {Sule}},
  \bibinfo {author} {\bibfnamefont {G.~Y.}\ \bibnamefont {Cho}}, \bibinfo
  {author} {\bibfnamefont {S.}~\bibnamefont {Ryu}}, \ and\ \bibinfo {author}
  {\bibfnamefont {R.}~\bibnamefont {Leigh}},\ }\href@noop {} {\  (\bibinfo
  {year} {2014})},\ \Eprint {http://arxiv.org/abs/arxiv:1403.6902}
  {arxiv:1403.6902} \BibitemShut {NoStop}%
\bibitem [{\citenamefont {Cappelli}\ and\ \citenamefont
  {Randellini}(2013)}]{Cappelli2013}%
  \BibitemOpen
  \bibfield  {author} {\bibinfo {author} {\bibfnamefont {A.}~\bibnamefont
  {Cappelli}}\ and\ \bibinfo {author} {\bibfnamefont {E.}~\bibnamefont
  {Randellini}},\ }\href {\doibase 10.1007/JHEP12(2013)101} {\bibfield
  {journal} {\bibinfo  {journal} {Journal of High Energy Physics}\ }\textbf
  {\bibinfo {volume} {2013}},\ \bibinfo {eid} {101} (\bibinfo {year} {2013}),\
  10.1007/JHEP12(2013)101}\BibitemShut {NoStop}%
\bibitem [{\citenamefont {Hsieh}\ \emph {et~al.}(2015)\citenamefont {Hsieh},
  \citenamefont {Cho},\ and\ \citenamefont {Ryu}}]{Shinsei2015}%
  \BibitemOpen
  \bibfield  {author} {\bibinfo {author} {\bibfnamefont {C.-T.}\ \bibnamefont
  {Hsieh}}, \bibinfo {author} {\bibfnamefont {G.~Y.}\ \bibnamefont {Cho}}, \
  and\ \bibinfo {author} {\bibfnamefont {S.}~\bibnamefont {Ryu}},\ }\href@noop
  {} {\  (\bibinfo {year} {2015})},\ \Eprint
  {http://arxiv.org/abs/arxiv:1403.6902} {arxiv:1403.6902} \BibitemShut
  {NoStop}%
\bibitem [{\citenamefont {Yao}\ and\ \citenamefont {Ryu}(2013)}]{Yao2013}%
  \BibitemOpen
  \bibfield  {author} {\bibinfo {author} {\bibfnamefont {H.}~\bibnamefont
  {Yao}}\ and\ \bibinfo {author} {\bibfnamefont {S.}~\bibnamefont {Ryu}},\
  }\href@noop {} {\bibfield  {journal} {\bibinfo  {journal} {Phys. Rev. B}\
  }\textbf {\bibinfo {volume} {88}},\ \bibinfo {pages} {064507} (\bibinfo
  {year} {2013})}\BibitemShut {NoStop}%
\bibitem [{\citenamefont {Polchinski}(1998)}]{Polchinski}%
  \BibitemOpen
  \bibfield  {author} {\bibinfo {author} {\bibfnamefont {J.}~\bibnamefont
  {Polchinski}},\ }\href@noop {} {\emph {\bibinfo {title} {String Theory}}}\
  (\bibinfo  {publisher} {Cambridge University Press},\ \bibinfo {year}
  {1998})\BibitemShut {NoStop}%
\bibitem [{\citenamefont {Witten}(1985)}]{Witten1985}%
  \BibitemOpen
  \bibfield  {author} {\bibinfo {author} {\bibfnamefont {E.}~\bibnamefont
  {Witten}},\ }\href@noop {} {\bibfield  {journal} {\bibinfo  {journal}
  {Commum. Math. Phys.}\ }\textbf {\bibinfo {volume} {100}},\ \bibinfo {pages}
  {197} (\bibinfo {year} {1985})}\BibitemShut {NoStop}%
\bibitem [{\citenamefont {Cappelli}\ and\ \citenamefont
  {Zemba}(1997)}]{Cappelli1997}%
  \BibitemOpen
  \bibfield  {author} {\bibinfo {author} {\bibfnamefont {A.}~\bibnamefont
  {Cappelli}}\ and\ \bibinfo {author} {\bibfnamefont {G.~R.}\ \bibnamefont
  {Zemba}},\ }\href@noop {} {\bibfield  {journal} {\bibinfo  {journal} {Nucl.
  Phys. B}\ }\textbf {\bibinfo {volume} {490}},\ \bibinfo {pages} {595}
  (\bibinfo {year} {1997})}\BibitemShut {NoStop}%
\bibitem [{\citenamefont {Hawking}(1977)}]{Hawking1977}%
  \BibitemOpen
  \bibfield  {author} {\bibinfo {author} {\bibfnamefont {S.}~\bibnamefont
  {Hawking}},\ }\href@noop {} {\bibfield  {journal} {\bibinfo  {journal}
  {Commun. Math. Phys.}\ }\textbf {\bibinfo {volume} {55}},\ \bibinfo {pages}
  {133} (\bibinfo {year} {1977})}\BibitemShut {NoStop}%
\bibitem [{\citenamefont {Actor}(1991)}]{Actor}%
  \BibitemOpen
  \bibfield  {author} {\bibinfo {author} {\bibfnamefont {A.}~\bibnamefont
  {Actor}},\ }\href@noop {} {\bibfield  {journal} {\bibinfo  {journal} {J.
  Phys. A}\ }\textbf {\bibinfo {volume} {24}},\ \bibinfo {pages} {3741}
  (\bibinfo {year} {1991})}\BibitemShut {NoStop}%
\bibitem [{\citenamefont {Elizalde}(2012)}]{Elizalde}%
  \BibitemOpen
  \bibfield  {author} {\bibinfo {author} {\bibfnamefont {E.}~\bibnamefont
  {Elizalde}},\ }\href@noop {} {\emph {\bibinfo {title} {Ten Physical
  Applications of Spectral Zeta Functions}}}\ (\bibinfo  {publisher}
  {Springer},\ \bibinfo {year} {2012})\BibitemShut {NoStop}%
\bibitem [{\citenamefont {Fujikawa}(1979)}]{Fujikawa1979}%
  \BibitemOpen
  \bibfield  {author} {\bibinfo {author} {\bibfnamefont {K.}~\bibnamefont
  {Fujikawa}},\ }\href@noop {} {\bibfield  {journal} {\bibinfo  {journal}
  {Phys. Rev. Lett.}\ }\textbf {\bibinfo {volume} {42}},\ \bibinfo {pages}
  {1195} (\bibinfo {year} {1979})}\BibitemShut {NoStop}%
\bibitem [{\citenamefont {Fujikawa}(1980)}]{Fujikawa1980}%
  \BibitemOpen
  \bibfield  {author} {\bibinfo {author} {\bibfnamefont {K.}~\bibnamefont
  {Fujikawa}},\ }\href@noop {} {\bibfield  {journal} {\bibinfo  {journal}
  {Phys. Rev. D}\ }\textbf {\bibinfo {volume} {21}},\ \bibinfo {pages} {2848}
  (\bibinfo {year} {1980})}\BibitemShut {NoStop}%
\bibitem [{\citenamefont {Nielsen}\ and\ \citenamefont
  {Ninomiya}(1981)}]{Nielsen1981}%
  \BibitemOpen
  \bibfield  {author} {\bibinfo {author} {\bibfnamefont {H.}~\bibnamefont
  {Nielsen}}\ and\ \bibinfo {author} {\bibfnamefont {M.}~\bibnamefont
  {Ninomiya}},\ }\href@noop {} {\bibfield  {journal} {\bibinfo  {journal}
  {Phys. Lett. B}\ }\textbf {\bibinfo {volume} {105}},\ \bibinfo {pages} {219}
  (\bibinfo {year} {1981})}\BibitemShut {NoStop}%
\bibitem [{\citenamefont {Nielsen}\ and\ \citenamefont
  {Ninomiya}(1983)}]{Nielsen1983}%
  \BibitemOpen
  \bibfield  {author} {\bibinfo {author} {\bibfnamefont {H.}~\bibnamefont
  {Nielsen}}\ and\ \bibinfo {author} {\bibfnamefont {M.}~\bibnamefont
  {Ninomiya}},\ }\href@noop {} {\bibfield  {journal} {\bibinfo  {journal}
  {Phys. Lett. B}\ }\textbf {\bibinfo {volume} {130}},\ \bibinfo {pages} {389}
  (\bibinfo {year} {1983})}\BibitemShut {NoStop}%
\bibitem [{\citenamefont {Coxeter}\ and\ \citenamefont
  {Moser}(1980)}]{groupbook}%
  \BibitemOpen
  \bibfield  {author} {\bibinfo {author} {\bibfnamefont {H.~S.~M.}\
  \bibnamefont {Coxeter}}\ and\ \bibinfo {author} {\bibfnamefont {W.~O.~J.}\
  \bibnamefont {Moser}},\ }\href@noop {} {\emph {\bibinfo {title} {Generators
  and Relations for Discrete Groups}}}\ (\bibinfo  {publisher} {Springer},\
  \bibinfo {year} {1980})\BibitemShut {NoStop}%
\end{thebibliography}%

\pagebreak

\onecolumngrid
\appendix

\setcounter{equation}{0}

\setcounter{figure}{0}

\setcounter{table}{0}

\setcounter{page}{1}

\makeatletter

\renewcommand{\theequation}{A\arabic{equation}}

\section{Appendix A: General Properties of the Modular Transformation}
In this section, we consider the general properties of the modular transformation, $A$, on the $d$-dimensional torus, $T^d$, written as
\bea
\label{Eq:Trperi}
\hat{L}'_{\mu}=A_{\mu\nu}\hat{L}_\nu,
\eea
where $A\in{}PSL(d,\mathbb{Z})$ and $\hat{L}_\mu$ is the period of the torus along the $\mu$-th direction. The period of the torus is a vector within the torus of which a separate coordinate vector, $\mathbf{x}$, satisfies
\bea
\mathbf{x}=\mathbf{x}+n_\mu\hat{L}_\mu
\eea
where $n_\mu$ is a integer. In $(1+1)$-D, we can express $\hat{L}_\mu$ utilizing simple complex coordinates, however, in $(2+1)$-D and $(3+1)$-D, we must express the period in quaternion coordinates. A vector, $\vec{x}$, on the torus may be parameterized using $x_{\mu}\in[0,1)$ as,
\bea
\vec{x}=x_\mu \hat{L}_\mu.
\eea
As we are interested in the modular transformations within $(2+1)$-D and $(3+1)$-D systems, it is important to identify the relevant generators within the relevant $PSL(d,\mathbb{Z})$ group. Therefore, we identify the generators of the group $PSL(3,\mathbb{Z})$ that are given by\cite{groupbook}
\bea
\label{Eq:Trmat}
S=\left( \begin{matrix} 0 & 1 & 0 \\ 0 & 0 & 1  \\ 1 & 0 & 0  \\ \end{matrix}\right)
,T=\left( \begin{matrix} 1 & 1 & 0 \\ 0 & 1 & 0  \\ 0 & 0 & 1  \\ \end{matrix}\right)
\eea
and the generators of the group $PSL(4,\mathbb{Z})$ are given by
\bea
\label{Eq:Trmat2}
S=\left( \begin{matrix} 0 & -1 & 0 & 0 \\ 0 & 0 & 1 & 0 \\ 0 & 0 & 0 & 1 \\ 1 & 0 & 0 & 0 \\ \end{matrix}\right)
,T=\left( \begin{matrix} 1 & 1 & 0 & 0 \\ 0 & 1 & 0 & 0 \\ 0 & 0 & 1 & 0 \\ 0 & 0 & 0 & 1 \\ \end{matrix}\right).
\eea
By applying the matrices shown in Eq. (\ref{Eq:Trmat}) and Eq. (\ref{Eq:Trmat2})  on Eq. $\eqref{Eq:Trperi}$, we see that the generators $S$ and $T$ act on the modular parameter in the same manner as that of the modular transformation in higher dimensions, namely: $S:(L_0,L_1,L_2,L_3)\rightarrow(-L_1,L_2,L_3,L_0)$ and $T:(L_0,L_1,L_2,L_3)\rightarrow(L_0+L_1,L_1,L_2,L_3)$.(In $(2+1)$-D, $S:(L_0,L_1,L_2)\rightarrow(L_1,L_2,L_0)$ and $T:(L_0,L_1,L_2)\rightarrow(L_0+L_1,L_1,L_2)$)

We now explore action of the modular transform by imposing that the coordinate vector $\mathbf{x}$ is invariant under the following transformation,
\bea
\mathbf{x}=x_\mu\hat{L}_\mu=x_\mu{A}^{-1}_{\mu\nu}\hat{L}'_\nu,
\eea
This allows one to show that the components of the coordinate vector, $\mathbf{x}$, transform as
\bea
x'_\mu=(A^{-1})^T_{\mu\nu}x_\nu.
\eea
Meanwhile, derivatives transform as,
\bea
\partial_\mu=\frac{\partial{x'}_\nu}{\partial{x}_\mu}\frac{\partial}{\partial'_\nu}=A^{-1}_{\mu\nu}\partial'_\nu.
\eea
and the various fermionic fields transform via,
\bea
\label{Eq:boundary0}
\psi'_{\lambda'}(x'):=\psi_\lambda(x).
\eea
It then follows that we may write the transformed fermionic fields as,
\bea
\label{Eq:boundary1}
\psi'_{\lambda'}(x)=\psi_\lambda({A}^T_{\mu\nu}x).
\eea
It is necessary to define the relevant boundary conditions for the corresponding fermionic fields. In $(2+1)$-D, the corresponding boundary conditions, $\vec{\lambda}_{2D}=(\lambda_0,\lambda_1,\lambda_2)$, and $(3+1)$-D, $\vec{\lambda}_{3D}=(\lambda_0,\lambda_1,\lambda_2,\lambda_3)$, may be defined as,
\bea
\label{Eq:boundary2}
\psi_\lambda(x_\nu\hat{L}_\nu+\hat{L}_\mu)=e^{2\pi i\lambda_\mu}\psi_\lambda(x_\nu\hat{L}_\nu).
\eea
By using Eq. \eqref{Eq:boundary0} and Eq. \eqref{Eq:boundary2}, we find the modular transformation of the boundary conditions to be,
\begin{gather}
\label{Eq:bc3}
e^{2\pi i \lambda_\mu}\psi_\lambda(x_\nu\hat{L}_\nu)=\psi_\lambda(x_\nu\hat{L}_\nu+\hat{L}_\mu)
\nonumber
\\={\psi'}_{\lambda'}(x'_\nu \hat{L'}_\nu+(A^{-1})^T_{\rho\mu}\hat{L'}_\rho)=e^{2\pi i \lambda'_\rho (A^{-1})^T_{\rho\mu}}{\psi'}_{\lambda'}(x'_\nu \hat{L'}).
\end{gather}
By equating the two phases on the left and right hand sides of Eq. \eqref{Eq:bc3}, the boundary conditions transform as,
\bea
\lambda'_\mu=A_{\mu\nu}\lambda_\nu.
\eea
Finally, we must consider the how the Lagrangian transforms under the modular transformation. This can be accomplished through,
\bea
\label{eq:action}
\mathcal{L}[\psi_\lambda(x),L_\mu^{-1}\partial_\mu\psi_\lambda(x)]= \mathcal{L}[\psi'_{\lambda'}(x'),L_\mu^{-1}A^{-1}_{\mu\nu}\partial'_\nu\psi'_{\lambda'}(x')].
\eea
\setcounter{equation}{0}
\renewcommand{\theequation}{B\arabic{equation}}
\section{Appendix B: Calculation of the Partition Function in $(1+1)$-D}
We begin the calculation of the partition function in $(1+1)$-D by writing the unregularized partition function of the $(1+1)$-D action.
For a chiral $(1+1)$-D edge with the dispersion $E(k)=k$, the temperature $\beta=\frac{1}{T}$ and the length $L$ in $x$ direction, we have the action that may be defined as,
\begin{gather}
S_{\lambda_0,\lambda_1}[\bar\psi_\lambda,\psi_\lambda]
\\
\nonumber
=-\beta{L}\int_0^1dx\int_0^1d\tau[\bar\psi(\tau,x)_\lambda(\partial_\tau/\beta+(-i\partial_x/L))\psi_\lambda(\tau,x)]
\\
\nonumber
=-\int_0^1dx\int_0^1d\tau\bar\psi(\tau,x)_\lambda(L\partial_\tau-i\beta\partial_x)\psi(\tau,x)_\lambda,
\end{gather}
where $\lambda_{0,1}$ denotes the boundary condition of the fermionic field $\psi_\lambda$ as either periodic or anti-periodic respectively. Under a modular transformation, $A_{\mu\nu}$, the action transforms as shown in Eq. \eqref{eq:action},
\bea
\nonumber
S_{\lambda_0,\lambda_1}[\bar\psi,\psi]&=&-\int_0^1dx\int_0^1d\tau\bar\psi_\lambda(\tau,x)[L(A_{22}\partial_\tau-A_{12}\partial_x)-i\beta(A_{11}\partial_x-A_{21}\partial_\tau)]\psi(\tau,x)_\lambda
\\
&=&-\int_0^1dx\int_0^1d\tau\bar\psi_\lambda(\tau,x)(L_1\partial_\tau-L_0\partial_x)\psi_\lambda(\tau,x),
\eea
where $L_{0}=A_{11}i\beta+A_{12}L$ and $L_{1}=A_{21}i\beta+A_{22}L$ are the periods of the torus, as defined in Appendix A. After the Fourier transformation along the time and the space directions, the action becomes
\bea
S_{\lambda_0,\lambda_1}[\bar a,a]=-\sum_{n_0,n_1}\bar a_{n_0,n_1}(2\pi L_1(n_0+\lambda_0)-2\pi L_0(n_1+\lambda_1))a_{n_0,n_1}.
\eea
GIven the action, it is then a simple task to find the corresponding partition function that is given as,
\begin{gather}
Z_{\lambda_0,\lambda_1}=\prod_{n_0,n_1=-\infty}^\infty\frac{1}{L_1} (2\pi i L_1(n_0+\lambda_0)-2\pi i \L_0(n_1+\lambda_1))
\\
\nonumber
=\prod_{n_0,n_1=-\infty}^\infty(2\pi i (n_0+\lambda_0)-2\pi i \tau(n_1+\lambda_1)),
\end{gather}
where $\tau=\frac{L_0}{L_1}$. The factor $\frac{1}{L_1}$ comes from the normalization factor of the path integral integrand. Using the Matsubara summation formula, we obtain a more simplified expression for the partition function,
\bea\label{eq:formalZ}
Z_{\lambda_0\lambda_1}=\prod_{n_1=-\infty}^{\infty}(1-e^{2i\pi{}\tau(n_1+\lambda_1)+2i\pi\lambda_0}).
\eea
 With the definition of the partition function, we now regularize the partition function as
\bea
\label{eq:formalZ}
Z_{\lambda_0\lambda_1}=\prod_{n_1=-\infty}^{\infty}(1-e^{2i\pi{}\tau(n_1+\lambda_1)+2i\pi\lambda_0})
\\
=\prod_{n_1=-\infty}^{-1}(1-e^{2i\pi{}\tau(n_1+\lambda_1)+2i\pi\lambda_0})\prod_{n_1=0}^{\infty}(1-e^{(2i\pi{}\tau(n_1+\lambda_1)+2i\pi\lambda_0)})
\\
=\prod_{n_1=-\infty}^{-1}-(1-e^{-2i\pi{}\tau(n_1+\lambda_1)-2i\pi\lambda_0})e^{2i\pi{}\tau(n_1+\lambda_1)+2i\pi\lambda_0}\prod_{n_1=0}^{\infty}(1-e^{2i\pi{}\tau(n_1+\lambda_1)+2i\pi\lambda_0})
\\=[\prod_{n_1=-\infty}^{-1}e^{-2\pi i(1/2-\lambda_0)}e^{2\pi i \tau(n_1+\lambda_1)}][(1-\omega)\prod_{n_1=1}^{\infty}(1-\omega q^{n_1})(1-\omega^{-1}q^{n_1})]
\\=[e^{-2\pi i\sum_{n_1=-\infty}^{-1}(1/2-\lambda_0)}e^{2\pi i\sum_{n_1=-\infty}^{-1} \tau(n_1+\lambda_1)}][(1-\omega)\prod_{n_1=1}^{\infty}(1-\omega q^{n_1})(1-\omega^{-1}q^{n_1})]
\\=[e^{-2\pi i(1/2-\lambda_0)\sum_{n_1=-\infty}^{-1}1}e^{2\pi i\tau\sum_{n_1=-\infty}^{-1} (n_1+\lambda_1)}][(1-\omega)\prod_{n_1=1}^{\infty}(1-\omega q^{n_1})(1-\omega^{-1}q^{n_1})],
\eea
where $q=e^{2\pi i \tau}$, $\omega=e^{2\pi i \lambda_0}q^{\lambda_1}$. In order to complete the proof, we use the Hurwitz zeta function that is defined as,
\bea
\zeta(-1,\lambda_1)=\sum_{n_1=0}^\infty (n_1+\lambda_1)= \frac{-1}{2}({\lambda_1}^2-\lambda_1+1/6)
\\
\zeta(0,\lambda_1)=\sum_{n_1=0}^\infty (n_1+\lambda_1)^0= (1/2-\lambda_1).
\eea
By using the following identities, we change the above summations from those over positive integers to a summation covering negative integers so as to complete the above summations by using following identities:
\bea
\sum_{n_1=-\infty}^{-1} (n_1+\lambda_1)=-\sum_{n_1=0}^{\infty} (n_1+1-\lambda_1)=-\zeta(-1,1-\lambda_1)=\frac{1}{2}({\lambda_1}^2-\lambda_1+1/6)
\eea
and,
\bea
\sum_{n_1=-\infty}^{-1} (n_1+\lambda_1)^0=\sum_{n_1=0}^{\infty} (n_1+1-\lambda_1)^0=\zeta(0,1-\lambda_1)=-(1/2-\lambda_1).
\eea
By substituting the above summations into the expression of the partition function found in \eqref{eq:formalZ}, the final form of the $(1+1)$-D regularized partition function is given by
\bea
Z=[e^{2\pi i(1/2-\lambda_0)(1/2-\lambda_1)}q^{(\lambda_1^2-\lambda_1+1/6)/2}][(1-\omega)\prod_{n_1=1}^{\infty}(1-\omega q^{n_1})(1-\omega^{-1}q^{n_1})].
\eea
\setcounter{equation}{0}
\renewcommand{\theequation}{C\arabic{equation}}
\section{Appendix C: Assigning the Determinant of the $C$ Matrix}
\subsection{1.$C$ matrix under the $T$ transformation}
We begin by reproducing the modular anomaly of $(1+1)$-D chiral edge. The basis which diagonalizes the transformation matrix, $C$, under the $T$ transformation is given according to Eq. (11) in the main text.
\begin{gather}
\ket{\theta,n_1}_{\lambda_0,\lambda_1}=\Phi_{\lambda_0,\lambda_1} \sum_{n_0=0}^{n_1-1}\sum_{j=-\infty}^{\infty} e^{2\pi i (\widetilde{n}_0+{n_1}j)\theta} T^j[\Phi_{n_0,n_1}],
\end{gather}
where $\theta\in[0,1)$. Application of the $T$ transformation to $\ket{\theta,n_1}_{\lambda_0,\lambda_1}$ is given as,
\bea
T\ket{\theta,n_1}_{\lambda_0,\lambda_1}=T[\Phi_{\lambda_0,\lambda_1} \sum_{n_0=0}^{n_1-1}\sum_{j=-\infty}^{\infty} e^{2\pi i (\widetilde{n}_0+{n_1}j)\theta} T^j[\Phi_{n_0,n_1}]]
\\
\nonumber
=T[ \sum_{n_0=0}^{n_1-1}\sum_{j=-\infty}^{\infty} e^{2\pi i (\widetilde{n}_0+{n_1}j)\theta}\Phi_{n_0+ j n_1+\lambda_0,n_1+\lambda_1}]
\\
\nonumber
= \sum_{n_0=0}^{n_1-1}\sum_{j=-\infty}^{\infty} e^{2\pi i (\widetilde{n}_0+{n_1}j)\theta}\Phi_{(n_0+n_1)+ j n_1+(\lambda_0+\lambda_1),n_1+\lambda_1}
\\
\nonumber
=e^{-2\pi i(n_1+\lambda_1) \theta} \Phi_{\lambda_0+\lambda_1,\lambda_1} \sum_{n_0=0}^{n_1-1}\sum_{j=-\infty}^{\infty} e^{2\pi i ((\widetilde{n}_0+\lambda_1)+{n_1}(j+1))\theta}T^{j+1}[\Phi_{n_0,n_1}]
\\
\nonumber
=e^{-2\pi i(n_1+\lambda_1) \theta}\ket{\theta,n_1}_{\lambda_0+\lambda_1,\lambda_1}.
\eea
As a result, the newly selected basis diagonalizes $C^{1D}_{T}$ matrix resulting in Eq. (13) of the main text, namely
\bea
C^{1D}_{T,\{\theta,n_1,\theta',n_1'\}}=(e^{-2\pi i (n_1+\lambda)\theta})\delta(\theta-\theta')\delta_{n_1,n_1'}.
\eea
After the diagonalization of $C^{1D}_{T}$ matrix, the determinant is given as the product of the diagonal entries. We divide the partition function of the path integral form into the anomalous divergent contribution, $Z_A$, and the regular contribution, $Z_R$, with the total partition function, $Z_{total}=Z_A Z_R$. In the calculation of the total partition function, we note that the anomalous contribution, $Z_A$, is the path integral of the negative dispersion modes only. The regular contribution to the total partition function, $Z_R$ is invariant under the $T$ transformation, $Z_{R,\vec{\lambda}}(\tau+1)=\prod_{n_1=0}^{\infty}(1-e^{(2i\pi{}\tau(n_1+\lambda_1)+2i\pi(\lambda_0+\lambda_1))})=Z_{R,\vec{\lambda}'}$ indicating that the contribution to the modular anomaly under the $T$ transformation comes entirely from $Z_A$. In other words, the regularized form of the total partition function transforms under the $T$ transformation as
\begin{gather}
Z_{total,\vec{\lambda}}(\tau+1)=[Z_{A,\vec{\lambda}}(\tau+1)][Z_{R,\vec{\lambda}}(\tau+1)]=[{C^{1D}_{T}}^{-2}Z_{A,\vec{\lambda}'}(\tau)][Z_{R,\vec{\lambda}'}(\tau)].
\\
\nonumber
\end{gather}
Therefore, we regularize the change of the measure of $Z_A$, which restricts the $C$ matrix to the negative momentums. Then, the anomalous phase of $C^{1D}_{T}$ is given by
\bea
\label{1DTsum}
arg(det(C^{1D}_{T}))= -2\pi[ \int_0^1 d\theta \theta][\sum_{n_1=-\infty}^{-1} (n_1+\lambda)]=-\pi\frac{1}{2}(\lambda^2-\lambda+1/6).
\eea
To extend the calculation beyond $(1+1)$-D, we use the matrices given by Eq. (13) and (16) of the main text to extend to $(2+1)$-D and $(3+1)$-D respectively. With the addition of the requisite extra dimensional momentum indices, we can write the phases of $det(C)$ in the same form as of Eq. (\ref{1DTsum}) for both $(2+1)$-D and $(3+1)$-D cases as,
\begin{gather}
arg(det(C^{2D}_{T}))= -2\pi  \sum_{n_1,n_2\in\mathbb{Z}} \int_0^1 d\theta \theta (n_1+\lambda_1)
\\
\nonumber
=-\pi  \sum_{n_1,n_2\in\mathbb{Z}} (n_1+\lambda_1)=-\sum_{n_1,n_2=-\infty}^\infty \frac{L_x^2}{8\pi}\frac{\partial}{\partial\lambda_1}(F_2)^2.
\end{gather}
and
\begin{gather}
arg(det(C^{3D}_{T}))= -2\pi  \sum_{n_1,n_2,n_3\in\mathbb{Z}} \int_0^1 d\theta \theta  (n_1+\lambda_1)
\\
\nonumber
=-\pi  \sum_{n_1,n_2,n_3\in\mathbb{Z}}  (n_1+\lambda_1)=-\sum_{n_1,n_2,n_3=-\infty}^\infty \frac{L_x^2}{8\pi}\frac{\partial}{\partial\lambda_1}(F_3)^2,
\end{gather}
where $F_2=2\pi\sqrt{\widetilde{n_1}^2/L_x^2+\widetilde{n_2}^2/L_y^2}$ is the dispersion in $(2+1)$-D and $F_3=2\pi\sqrt{\widetilde{n_1}^2/L_x^2+\widetilde{n_2}^2/L_y^2+\widetilde{n_3}^2/L_z^2}$ is the dispersion in $(3+1)$-D. In order to evaluate the above summations, we must define the following Epstein-Zeta functions in which we use the variable $\epsilon$ to denote the scale that cuts off the high energy states:
\bea
E_2(\epsilon,c_1,c_2,a_1,a_2)\equiv\sum_{n_1,n_2=0}^\infty(a_1(n_1+c_1)^2+a_2(n_2+c_2)^2)^{-\epsilon},
\eea
\bea
G_2(\epsilon,c_1,c_2,a_1,a_2)\equiv\sum_{n_1,n_2=-\infty}^\infty(a_1(n_1+c_1)^2+a_2(n_2+c_2)^2)^{-\epsilon},
\eea
\begin{gather}
E_3(\epsilon,c_1,c_2,c_3,a_1,a_2,a_3)\equiv
\\
\nonumber
\sum_{n_1,n_2,n_3=0}^\infty(a_1(n_1+c_1)^2+a_2(n_2+c_2)^2+a_3(n_3+c_3)^2)^{-\epsilon},
\end{gather}
\begin{gather}
G_3(\epsilon,c_1,c_2,c_3,a_1,a_2,a_3)\equiv
\\
\nonumber
\sum_{n_1,n_2,n_3=-\infty}^\infty(a_1(n_1+c_1)^2+a_2(n_2+c_2)^2+a_3(n_3+c_3)^2)^{-\epsilon},
\end{gather}
\begin{gather}
G_3(\epsilon,c_1,c_2,c_3,a_1,a_2,a_3)\equiv
\\
\nonumber
\sum_{n_1,n_2=0,n_3=-\infty}^\infty(a_1(n_1+c_1)^2+a_2(n_2+c_2)^2+a_3(n_3+c_3)^2)^{-\epsilon}.
\end{gather}
We substitute previous summations over the dispersion relations, $F$, into the newly defined Zeta function expressions to obtain the anomalous phase resulting from the application of the $T$-transform in higher dimensions as
\bea
\label{ct2d}
arg(det(C_T^{2D}))=-\frac{\pi}{2}L_x^2\frac{\partial}{\partial\lambda_1}G_2(-1,\lambda_1,\lambda_2,1/L_x^2,1/L_y^2)
\eea
in $(2+1)$-D and
\bea
\label{ct3d}
arg(det(C_T^{3D}))=-\frac{\pi}{2}L_x^2\frac{\partial}{\partial\lambda_1}G_3(-1,\lambda_1,\lambda_2,\lambda_3,1/L_x^2,1/L_y^2,1/L_z^2)
\eea
in $(3+1)$-D. In fact, in order to perform the divergent summations that arise within both the $S$ and $T$ transformations in the $(2+1)$-D and $(3+1)$-D cases, we use following recursion equation\cite{Elizalde},
\begin{gather}
\label{recurs}
E_N(\epsilon,c_1...c_N,a_1,..a,_N)=\frac{1}{\Gamma(\epsilon)}\sum_{m=0}^\infty \frac{(-1)^m}{m!}a_1^m\zeta(-2m,c_1)\Gamma(\epsilon+m)E_{N-1}(\epsilon+m,c_2...c_N,a_2,..a_N)
\\+\frac{1}{2}\sqrt{\frac{\pi}{a_1}}\frac{\Gamma(\epsilon-\frac{1}{2})}{\Gamma(\epsilon)}E_{N-1}(\epsilon-\frac{1}{2},c_2,...c_N,a_2,...,a_N)
\nonumber
\\+\sqrt{\frac{\pi}{a_1}}\frac{cos(2\pi c_1)}{\Gamma(\epsilon)}\sum_{n_1=1}^\infty\sum_{n_2,n_3..n_N=0}^\infty \int_0^\infty t^{\epsilon-3/2}exp[-\frac{\pi^2n_1^2}{a_1t}-t\sum_{j=1}^{N}a_j^2(n_j+c_j)^2] dt.
\nonumber
\end{gather}
In \eqref{recurs}, the gamma function $\Gamma(n)$ is given by $\Gamma(n)=(n-1)!$ when $n$ is positive integer and the function is divergent at $n=0,-1,-2..$. We additionally make use of the the gamma function identity $\Gamma(x+1)=x\Gamma(x)$ when evaluating the recursion relationship. In order to make \eqref{recurs} more tractable, we break the right hand side of the equation into distinct terms and label each of the terms as $E_A$, $E_B$, and $E_C$. The first of these terms, $A$, is given by
\begin{gather}
\label{eq:adef}
\nonumber
E_A=\zeta(0,c_1)E_{N-1}(\epsilon+0,c_2,..c_N,a_2,...,a_N)-\frac{\Gamma(\epsilon+1)}{\Gamma(\epsilon)}a_1\zeta(-2,c_1)E_{N-1}(\epsilon+1,c_2...c_N,a_2,...,a_N)+
\\
+\frac{1}{\Gamma(\epsilon)}\sum_{m=2}^\infty \frac{(-1)^m \Gamma(\epsilon+m)}{m!}a_1^m\zeta(-2m,c_1)E_{N-1}(\epsilon+m,c_2...c_N,a_2,...,a_N),
\end{gather}
where we have expanded out the $m=0$ and $m=1$ terms for clarity. We notice that we may immediately eliminate the last term in the \eqref{eq:adef} due to the divergent denominator. The second of these terms is $E_B$ that is defined as
\begin{align}
\label{eq:bdef}
\nonumber
E_B=\frac{1}{2}\sqrt{\frac{\pi}{a_1}}\frac{\Gamma(\epsilon-\frac{1}{2})}{\Gamma(\epsilon)}E_{N-1}(\epsilon-\frac{1}{2},c_2,...c_N,a_2,...,a_N).
\end{align}
It is clear that $E_B$ vanishes by inspection as the denominator is divergent while the numerator remains finite. The integral expression of $E_C$ can be rewritten from its form in \eqref{recurs} to a more simplified form as,
\begin{align}
E_C=\int_0^\infty dt t^{\epsilon-3/2}exp[-\frac{\pi^2n_1^2}{a_1t}-t\sum_{j=1}^{N}a_j^2(n_j+c_j)^2]
\\=\sqrt{\pi}\frac{e^{-2\sqrt{\frac{\pi^2n_1^2}{a_1}\sum_{j=1}^{N}a_j^2(n_j+c_j)^2}}}{\sqrt{\frac{\pi^2n_1^2}{a_1}}}.
\end{align}
when $\epsilon=0$ and in which $E_C$ becomes
\bea
= \sqrt{\pi }\frac{\left(1+2 \sqrt{\frac{\pi^2n_1^2}{a_1}} \sqrt{\sum_{j=1}^{N}a_j^2(n_j+c_j)^2}\right) e^{-2 \sqrt{\frac{\pi^2n_1^2}{a_1}} \sqrt{\sum_{j=1}^{N}a_j^2(n_j+c_j)^2}}}{2 (\frac{\pi^2n_1^2}{a_1})^{3/2}},
\eea
when $\epsilon=-1$. In the previous expressions for $E_C$ we observe that the integral expressions are exponentially decaying, therefore the entire numerator of $C$ is finite thereby resulting in a vanishing contribution from $E_C$.

Therefore, we conclude that when $\epsilon=-1,0$,
\begin{gather}
\label{finalrecurs}
E_N(\epsilon,c_1...c_N,a_1,..,a_N)=A
\\
\nonumber
=\zeta(0,c_1)E_{N-1}(\epsilon+0,c_2,...c_N,a_2,...,a_N)-\frac{\Gamma(\epsilon+1)}{\Gamma(\epsilon)}a_1\zeta(-2,c_1)E_{N-1}(\epsilon+1,c_2...c_N,a_2,...,a_N).
\end{gather}

The final recursion relation in Eq. (\ref{finalrecurs}) is evaluated for the $(2+1)$-D case ($N=2$) at $\epsilon=-1$ and $\epsilon=0$ in conjunction with the expression $E_{1}(\epsilon,c_1,a_1)=a_1^{-\epsilon}\zeta(2\epsilon,c_1)$, taken from its definition, to find that for the $(2+1)$-D case,
\begin{gather}
\nonumber
E_{2}(-1,\lambda_1,\lambda_2,a_1,a_2)=a_2\zeta(0,\lambda_1)\zeta(-2,\lambda_2)+a_1\zeta(-2,\lambda_1)\zeta(0,\lambda_2)
\\
\nonumber
=\frac{a_2}{3}(\lambda_1-1/2)(\lambda_2^3-\frac{3}{2}\lambda_2^2+\frac{1}{2}\lambda_2)+\frac{a_1}{3}(\lambda_2-1/2)(\lambda_1^3-\frac{3}{2}\lambda_1^2+\frac{1}{2}\lambda_1)
\\
\label{e2}
=-\frac{1}{3}(\lambda_1-\frac{1}{2})(\lambda_2-\frac{1}{2})[a_1\lambda_1(1-\lambda_1)+a_2\lambda_2(1-\lambda_2)],
\end{gather}
\begin{gather}
\label{Sres2D}
E_2(0,\lambda_1,\lambda_2,a_1,a_2)=\zeta(0,\lambda_1)\zeta(0,\lambda_2)=(1/2-\lambda_1)(1/2-\lambda_2).
\end{gather}
For the $(3+1)$-D case, we substitute in the expression of $E_2$ to the recursion relation of Eq. (\ref{finalrecurs}) to find that
\begin{gather}
\nonumber
E_3(-1,\lambda_1,\lambda_2,\lambda_3,a_1,a_2,a_3)
\\
\nonumber
=\zeta(0,\lambda_1)E_{2}(-1,\lambda_2,\lambda_3,a_2,a_3)+a_1\zeta(-2,\lambda_1)E_{2}(0,\lambda_2,\lambda_3,a_2,a_3)
\\
\nonumber
=\frac{1}{3}(\lambda_1-\frac{1}{2})(\lambda_2-\frac{1}{2})(\lambda_3-\frac{1}{2})[a_2\lambda_2(1-\lambda_2)+a_3\lambda_3(1-\lambda_3)]
\\
\nonumber-a_1\frac{1}{3}\lambda_1(\lambda_1-1)(\lambda_1-1/2)(1/2-\lambda_2)(1/2-\lambda_3)
\\
\label{e3}
=\frac{1}{3}(\lambda_1-\frac{1}{2})(\lambda_2-\frac{1}{2})(\lambda_3-\frac{1}{2})[a_1\lambda_1(1-\lambda_1)+a_2\lambda_2(1-\lambda_2)+a_3\lambda_3(1-\lambda_3)],
\end{gather}
\begin{gather}
\label{Sres3D}
E_3(0,\lambda_1,\lambda_2,\lambda_3,a_1,a_2,a_3)
\\
\nonumber
=\zeta(0,\lambda_1)E_{2}(0,\lambda_2,\lambda_3,a_2,a_3)=(1/2-\lambda_1)(1/2-\lambda_2)(1/2-\lambda_3).
\end{gather}

To complete the calculation of the $C$ matrix under $T$-transform, we must calculate $G$, corresponding to the summation over all complex numbers $\mathbb{Z}$, from the expression of the recursion relation, $E$, as the zeta function contained within $E$ is defined as the summation over positive integers. To accomplish this, we use the following identity\cite{Elizalde},
\begin{gather}
\label{Eq:sumc}
G_1(\epsilon,\lambda)\equiv\sum_{n=-\infty}^{n=\infty}((n+\lambda)^2)^{-\epsilon}=
(\sum_{n=-\infty}^{n=-1}+\sum_{n=0}^{n=\infty})((n+\lambda)^2)^{-\epsilon}
\\
\nonumber
=\sum_{n=0}^{n=\infty}((n+1-\lambda)^2)^{-\epsilon}+\sum_{n=0}^{n=\infty}((n+\lambda)^2)^{-\epsilon}=E_1(\epsilon,\lambda)+E_1(\epsilon,1-\lambda).
\end{gather}
When considering the $(2+1)$-D version of Eq. \eqref{Eq:sumc}, we learn that
\bea
G_2(\epsilon,\lambda_1,\lambda_2)=E_2(\epsilon,\lambda_1,\lambda_2)+E_2(\epsilon,1-\lambda_1,\lambda_2)
\\+E_2(\epsilon,\lambda_1,1-\lambda_2)+E_2(\epsilon,1-\lambda_1,1-\lambda_2).
\nonumber
\eea
Finally, for the $(3+1)$-D version of Eq. \eqref{Eq:sumc}, the result may be written as
\bea
G_3(\epsilon,\lambda_1,\lambda_2,\lambda_3)=E_3(\epsilon,\lambda_1,\lambda_2,\lambda_3)+E_3(\epsilon,1-\lambda_1,\lambda_2,\lambda_3)
\\+E_3(\epsilon,\lambda_1,1-\lambda_2,\lambda_3)+E_3(\epsilon,\lambda_1,\lambda_2,1-\lambda_3)
\nonumber
\\+E_3(\epsilon,\lambda_1,1-\lambda_2,1-\lambda_3)+E_3(\epsilon,1-\lambda_1,\lambda_2,1-\lambda_3)
\nonumber
\\+E_3(\epsilon,1-\lambda_1,1-\lambda_2,\lambda_3)+E_3(\epsilon,1-\lambda_1,1-\lambda_2,1-\lambda_3)
\nonumber
\eea
Nonetheless, when $\epsilon=0$ or $\epsilon=-1$, then the calculated expressions of the Epstein-Zeta functions in Eq. \eqref{e2} and Eq. \eqref{e3}, namely $E_{2}$ and $E_{3}$, may be simplified to satisfy
\bea
E_2(\epsilon,\lambda_1,\lambda_2)=-E_2(\epsilon,1-\lambda_1,\lambda_2)=-E_2(\epsilon,\lambda_1,1-\lambda_2),
\eea
and
\bea
E_3(\epsilon,\lambda_1,\lambda_2,\lambda_3)=-E_3(\epsilon,1-\lambda_1,\lambda_2,\lambda_3)
\\
\nonumber
=-E_3(\epsilon,\lambda_1,1-\lambda_2,\lambda_3)=-E_3(\epsilon,\lambda_1,\lambda_2,1-\lambda_3),
\eea
which yields the final sum over the complex numbers
\bea
\label{eq:gfinal}
G_2(\epsilon,\lambda_1,\lambda_2)=G_3(\epsilon,\lambda_1,\lambda_2,\lambda_3)=0.
\eea
Finally, using \eqref{eq:gfinal} and substituting the result into Eq. \eqref{ct2d} and Eq.  \eqref{ct3d}, we find that the contributions to the modular anomaly from the $T$ modular transformation in $(2+1)$-D is
\bea
arg(det(C_T^{2D}))=0,
\eea
meanwhile, in $(3+1)$-D, the contribution is
\bea
arg(det(C_T^{3D}))=0.
\eea
\subsection{2.$C$ matrix under the $S$ transformation}
We again start from the modular anomaly of a $(1+1)$-D edge. According to Eq. (11) in the main text, the basis that diagonalizes the $C$ matrix is given by
\begin{gather}
\ket{\phi,n_0,n_1}_{\lambda_0,\lambda_1}=\Phi_{\lambda_0-1/2,\lambda_1-1/2}(\Phi_{n_0+1/2,n_1+1/2}+e^{2\pi i \phi/4}\Phi_{-n_1+1/2,n_0+1/2}
\\
\nonumber+e^{2\pi i 2\phi/4}\Phi_{-n_0+1/2,-n_1+1/2}+e^{2\pi i 3\phi/4}\Phi_{n_1+1/2,-n_0+1/2})
\end{gather}
Where $\phi\in \lbrace -1,0,1,2\rbrace$. The application of the $S$ transformation to $\ket{\phi,n_0,n_1}$ is given as,
\begin{gather}
S\ket{\phi,n_0,n_1}_{\lambda_0,\lambda_1}
\\
\nonumber
=S(\Phi_{n_0+\lambda_0,n_1+\lambda_1}+e^{2\pi i \phi/4}\Phi_{-n_1+\lambda_0,n_0+\lambda_1}+e^{2\pi i 2\phi/4}\Phi_{-n_0+\lambda_0,-n_1+\lambda_1}+e^{2\pi i 3\phi/4}\Phi_{n_1+\lambda_0,-n_0+\lambda_1})
\\
\nonumber
=(\Phi_{-n_1-\lambda_1,n_0+\lambda_0}+e^{2\pi i \phi/4}\Phi_{-n_0-\lambda_1,-n_1+\lambda_0}+e^{2\pi i 2\phi/4}\Phi_{n_1-\lambda_1,-n_0+\lambda_0}+e^{2\pi i 3\phi/4}\Phi_{n_0-\lambda_1,n_1+\lambda_0})
\\
\nonumber
=\Phi_{-\lambda_1,\lambda_0}(\Phi_{-n_1,n_0}+e^{2\pi i \phi/4}\Phi_{-n_0,-n_1}+e^{2\pi i 2\phi/4}\Phi_{n_1,-n_0}+e^{2\pi i 3\phi/4}\Phi_{n_0,n_1})
\\
\nonumber
=e^{-2\pi i \phi/4}\Phi_{-\lambda_1,\lambda_0}(e^{2\pi i \phi/4}\Phi_{-n_1,n_0}+e^{2\pi i 2\phi/4}\Phi_{-n_0,-n_1}+e^{2\pi i 3\phi/4}\Phi_{n_1,-n_0}+\Phi_{n_0,n_1})
\\
\nonumber
=e^{-2\pi i \phi/4}\ket{\phi,n_0,n_1}_{-\lambda_1,\lambda_0}
\end{gather}

The $C$ matrix is then a diagonal matrix given by the expression,
\bea
C^{1D}_{S,\{\phi,n_0,n_1,\phi',n_0',n_1'\}}=(e^{-2\pi i \phi/N})\delta_{\phi,\phi'}\delta_{n_0,n_0'}\delta_{n_1,n_1'}.
\eea
As the determinant of diagonal matrix is the product of the diagonal entries, we have the unregulated phase of the $C$ matrix under the $S$-transform
\bea
arg(det(C^{1D}_{S}))= -2\pi [ \sum_{\phi=-1,0,1,2} \phi/4][\sum_{n_0=0}^\infty 1][\sum_{n_1=0}^\infty 1]
\eea
We regularize the above sum by attaching the following regulator.
\bea
-2\pi [ \sum_{\phi=-1,0,1,2} \phi/4] [\sum_{n_0=0}^\infty (n_0+\lambda_0)^0][\sum_{n_1=0}^\infty(n_1+\lambda_1)^0]
\\
\nonumber
=-\pi\zeta(0,\lambda_0) \zeta(0,\lambda_1)=- \pi  (1/2-\lambda_0)(1/2-\lambda_1)
\eea
To extend the calculation into higher dimensions, we use the matrices given by Eq. (13) and (16) of the main text. Using these, we write the phase of $det(C_{S})$ in $(2+1)$-D and $(3+1)$-D as,
\bea
arg(det(C^{2D}_{S}))= -4\pi [ \sum_{\phi=-1,0,1} \phi/3 ][\sum_{n_0\geq n_1 \geq n_2} 1]
\eea
\bea
arg(det(C^{3D}_{S}))= -4\pi  [\sum_{\phi=-3,-2,..,4} \phi/8][\sum_{n_0,n_1,n_2=0,n_3=-\infty}^\infty 1]
\eea
Without requiring the complete summation of the modes, we immediately see that $arg(det(C^{2D}_{S}))=0$ from the summation of $\phi$. To calculate $C^{3D}_{S}$, we again use the EZ zeta function recursion relation calculated in Eq. (\ref{finalrecurs}),
\bea
\label{connection2}
arg(det(C_S^{3D}))=-2\pi\sum_{n_0=0}^\infty \sum_{n_1,n_2=0,n_3=-\infty}^\infty (n_0+\lambda_0)^0 (F_3)^0,
\eea
The connection between the determinant of the transformation matrix, $det(C_{S}^{3D})$, and the EZ zeta function is given by
\bea
\label{connection3}
arg(det(C_S^{3D}))=-2\pi(1/2-\lambda_0) g_3(0,\lambda_1,\lambda_2,\lambda_3,a_1,a_2,a_3),
\eea
We derive the expression of $g_3$ from the expansion method used in Eq. \eqref{Eq:sumc}, we find
\begin{gather}
g_3(0,\lambda_1,\lambda_2,\lambda_3,a_1,a_2,a_3)
\\
\nonumber
=E_3(0,\lambda_1,\lambda_2,\lambda_3,a_1,a_2,a_3)+E_3(0,1-\lambda_1,\lambda_2,\lambda_3,a_1,a_2,a_3)
\\
\nonumber
+E_3(0,\lambda_1,1-\lambda_2,\lambda_3,a_1,a_2,a_3)+E_3(0,1-\lambda_1,1-\lambda_2,\lambda_3,a_1,a_2,a_3)=0.
\end{gather}
As a result, the contribution of  \eqref{connection3} vanishes and we conclude that the contributions to the modular anomaly resulting from the $S$ transformation in $(2+1)$-D to be
\bea
arg(det(C_S^{2D}))=0,
\eea
and in $(3+1)$-D
\bea
arg(det(C_S^{3D}))=0.
\eea
\setcounter{equation}{0}
\renewcommand{\theequation}{D\arabic{equation}}
\section{Appendix D: Numerical Calculation of the Casimir Energy}
Within this section, we introduce a numerical regularization scheme to calculate the Casimir energy in order to place our analytical regularization contained within the main text on a solid foundation. We start from the action of $(2+1)$-D theory given as
\bea
S_{\lambda_0\lambda_1\lambda_2}[\bar\psi,\psi]&=&-\beta{L_x}L_y\int_0^1\int_0^1\int_0^1dxdyd\tau\bar\psi_\lambda[\sigma_0\partial_\tau/\beta-i\sigma_x\partial_x/L_x-i\sigma_y\partial_y/L_y]\psi_\lambda
\\
\nonumber
&=&-\int_0^1\int_0^1\int_0^1dxdyd\tau\bar\psi_{\lambda}(\sigma_0L_xL_y,i\sigma_x\beta{L}_y,i\sigma_y\beta{L_x})(\partial_\tau,-\partial_x,-\partial_y)^T\psi_{\lambda}.
\eea
Under a modular transform $A_{\mu\nu}$ defined in Eq. \eqref{eq:action}, we have
\begin{gather}
\label{eq:2DZ}
S_{\lambda_0\lambda_1\lambda_2}[\bar\psi_\lambda,\psi_\lambda]
\\
\nonumber
=-\int_0^1\int_0^1\int_0^1dxdyd\tau\bar\psi_{\lambda}(\sigma_0L_xL_y,i\sigma_x\beta{L}_y,i\sigma_y\beta{L_x})A^{-1}(\partial_\tau,-\partial_x,-\partial_y)^T\psi_{\lambda}.
\\
\nonumber
=-\int_0^1\int_0^1\int_0^1dxdyd\tau\bar\psi_{\lambda}(L_3\partial_\tau-L_1\partial_x-L_2\partial_y)\psi_{\lambda},
\end{gather}
%where
%\bea
%L_1&=&(A_{22}A_{31}-A_{21}A_{32})\sigma_0L_xL_y+i(A_{22}A_{33}-A_{23}A_{32})\sigma_x\beta{L}_y+i(A_{23}A_{31}-A_{21}A_{33})\sigma_y\beta{L}_x,\\
%\nonumber
%L_2&=&(A_{11}A_{32}-A_{12}A_{31})\sigma_0L_xL_y+i(A_{13}A_{32}-A_{12}A_{33})\sigma_x\beta{L}_y+i(A_{11}A_{33}-A_{13}A_{31})\sigma_y\beta{L}_x,\\
%\nonumber
%L_3&=&(A_{11}A_{22}-A_{12}A_{21})\sigma_0L_xL_y+i(A_{13}A_{22}-A_{12}A_{23})\sigma_x\beta{L}_y+i(A_{11}A_{23}-A_{13}A_{21})\sigma_y\beta{L}_x.
%\eea
The partition function of Eq. (\ref{eq:2DZ}), after having been diagonalized, is given as
\begin{gather}
Z_{\lambda_0\lambda_1\lambda_2}
\\
\nonumber
=\prod_{n_0,n_1,n_2=-\infty}^{\infty}\det[\frac{1}{L_3}(2\pi i L_3 (n_0+\lambda_0)-2\pi iL_1(n_1+\lambda_1)-2\pi iL_2(n_2+\lambda_2))]
\\
\nonumber
=\prod_{n_1,n_2=-\infty}^{\infty}\det[I_{2}-\exp(i2\pi[z_1(n_1+\lambda_1)+z_2(n_2+\lambda_2)]+\lambda_0)]
\\
\nonumber
=\prod_{n_1,n_2=-\infty}^{\infty}(1-{e}^{2i\pi(z_{10}\widetilde{n}_1+z_{20}\widetilde{n}_2+\lambda_0)+2\pi{E}({n_1,n_2})})(1-{e}^{2i\pi(z_{10}\widetilde{n}_1+z_{20}\widetilde{n}_2+\lambda_0)-2\pi{E}({n_1,n_2})}),
\end{gather}
where $z_{i}\equiv z_{i0}I_{2} + i\sum_{j=1}^{2}z_{ij}\sigma_j \equiv\frac{L_{i}}{L_3}$, $\widetilde{n}_i=n_i+\lambda_i$ and ${E_2}(n_1,n_2)=\sqrt{\sum_{i=1,2,3}[z_{1i}\widetilde{n}_1+z_{2i}\widetilde{n}_2]^2}$, and $I_2$ is the $2x2$ identity matrix.

In $(3+1)$-D, we just add $z$-directional momentum. Similarly, we have
\begin{gather}
Z_{\lambda_0,\lambda_1,\lambda_2,\lambda_3}=\prod_{n_1,n_2,n_3=-\infty}^{\infty}(1-{e}^{2i\pi(\lambda_0+z_{10}\widetilde{n}_1+z_{20}\widetilde{n}_2+z_{30}\widetilde{n}_3)+2\pi{E_3}(n_1,n_2,n_3)})
\\
\nonumber
\times(1-{e}^{2i\pi(\lambda_0+z_{10}\widetilde{n}_1+z_{20}\widetilde{n}_2+z_{30}\widetilde{n}_3)-2\pi{E_3}(n_1,n_2,n_3)})
\end{gather}
where ${E_3}(n_1,n_2,n_3)=\sqrt{\sum_{i=1,2,3}[z_{1i}\widetilde{n}_1+z_{2i}\widetilde{n}_2+z_{3i}\widetilde{n}_3]^2}$.

In $(2+1)$-D and $(3+1)$-D, we extract the divergent part of the partition function so that we may use numerical regularization to find the anomalous contribution. In $(2+1)$-D, the divergent part is represented as
\bea
F^A_2&=&\sum_{n_1,n_2}2i\pi(z_{10}n_1+z_{20}n_2)+2\pi{E_2}(n_1,n_2)
\eea
and, in $(3+1)$-D, the contribution is
\bea
F^A_3&=&\sum_{n_1,n_2,n_3}2i\pi(z_{10}n_1+z_{20}n_2+z_{30}n_3)+2\pi{E_3}(n_1,n_2,n_3).
\eea
In general, we can evaluate the sum of a divergent function, $F^A=\sum_{\vec{m}}f_A(\vec{m})$, using following identity
\bea
\label{method}
F^A=\lim_{\epsilon\rightarrow 0} -\frac{\partial}{\partial\epsilon}\sum_{\vec{m}}e^{-f_A \epsilon},
\eea
that may be numerically evaluated using $\sum_{\vec{m}}e^{-f_A(\vec{m}) \epsilon}$. Therefore, from the numerical regularization, we extract the value of $O(\epsilon)$ by fitting the curve as a function of $\epsilon$ to evaluate $F_A$. From Eq. (\ref{method}), we find that the coefficient of $O(\epsilon)$ is the regularized Casimir energy. Nonetheless, as this quantity diverges when $\epsilon\rightarrow0$, it has poles of different orders in the expansion of $1/\epsilon$. For the sake of numerical stability, we need to subtract the divergent part of the partition function by calculating the analytical form of the poles. To calculate the divergent part analytically, we use the Euler-Maclaurin formula to change the sum $\sum_{\vec{m}}e^{-f_A \epsilon}$ to a integral. The Euler-Maclaurin formula states that we can express a sum, $\sum_{i=m+1}^n f(i)$, by
\bea
\sum_{i=m+1}^n f(i)=\int_m^n dx f(x)+B_1(f(n)-f(m))
\\
\nonumber
+\sum_{k=1}^p \frac{B_{2k}}{(2k)!}(f^{(2k-1)}(n)-f^{(2k-1)}(m))+R.
\eea
where $B_n$ is the $n$-th Bernoulli number($B_1=1/2$) and $R$ is an error term which becomes smaller in higher $p$-th order approximations. Consider $(2+1)$-D, all other terms except the integral can contribute the positive orders of $\epsilon$. The pole only comes from the integral.
\bea
\sum_{n_1,n_2=-\infty}^{\infty} e^{-2\pi (i (z_{10}n_1+z_{20}n_2)+E_2(n_1,n_2))\epsilon}
\\
\nonumber
=\sum_{n_2=-\infty}^\infty\int_{-\infty}^{\infty}dx e^{-2\pi (i (z_{10}x+z_{20}n_2)+E_2(x,n_2))\epsilon}
\\
\nonumber
=\int_{-\infty}^{\infty}dxdy e^{-2\pi (i (z_{10}x+z_{20}y)+E_2(x,y))\epsilon}
\\
\nonumber
=\frac{1}{\epsilon^2}\int_{-\infty}^{\infty}dxdy e^{-2\pi (i (z_{10}x+z_{20}y)+E_2(x,y))}.
\eea
Finally, in $(2+1)$-D, the integral expression which contributes to the pole is given by
\bea
S_{2D}=\frac{1}{\epsilon^2}\int^\infty_{-\infty}\int^\infty_{-\infty} dxdy e^{-2\pi(E_2(x,y)+i(z_{10}x+z_{20}y))}.
\eea
The above numerical regularization procedure may then be repeated with by adding in an additional coordinate so that, in $(3+1)$-D, we have
\bea
S_{3D}=\frac{1}{\epsilon^3}\int^\infty_{-\infty}\int^\infty_{-\infty}\int^\infty_{-\infty} dxdydz e^{-2\pi(E_3(x,y,z)+i(z_{10}x+z_{20}y+z_{30}z))}.
\eea
We calculate the integral part to evaluate the pole. In general, the expression for the integral in $(d+1)$-D is given by
\bea
\frac{1}{\epsilon^d} \int^\infty_{-\infty} d^dx e^{-2\pi(E_d(\vec{x})+i\sum_{j=1}^d z_{j0} x_j)}
\\
\nonumber
=\frac{1}{\epsilon^d}[\frac{1}{\sqrt{det(V)}} \int^\infty_{-\infty} d^dx' e^{-2\pi(\sum_{i=1}^d (x'_i)^2+i\sum_{j,k=1}^d z_{j0} w_{kj}x'_k)}],
\eea
where $x'_i=\sum_{j=1}^d z_{ji}x_j$ and $w_{ij}$ is the matrix component of the inverse matrix of $z_{ij}$ when $i,j$ are positive integers. We change to the polar coordinate to solve the above integral expression.
\bea
\\=\frac{1}{\epsilon^d\sqrt{det(V)}} \int^\infty_0 dr \int^\pi_0 d\theta S_{d-2}r^{d-1} sin^{d-2}\theta e^{-2\pi r(1+i|\sum_{j,k,l}z_{j0} w_{kj} w_{kl} z_{l0}|cos\theta) }
\\=\frac{1}{\epsilon^d\sqrt{det(V)}} \int^\infty_0 dr \int^\pi_0 d\theta S_{d-2}r^{d-1} sin^{d-2}\theta e^{-2\pi r(1+i\sqrt{b^T V^{-1} b}cos\theta) }
\\=\frac{1}{(2\pi\epsilon)^d\sqrt{det(V)}} S_{d-2}\Gamma(d) \int^\pi_0 d\theta \frac{sin^{d-2} \theta}{(1+i\sqrt{b^T V^{-1} b}cos\theta)^d}
\\=\frac{1}{(2\pi\epsilon)^d\sqrt{det(V)}} S_{d-2}\Gamma(d)  \int^1_{-1} dx \frac{(1-x^2)^{(d-3)/2}}{{(1+i\sqrt{b^T V^{-1} b}x)}^d},
\eea
where $V_{ij}=\sum_{k=1}^d z_{i k} z_{j k}$ and $b_i=z_{i0}$. When $d=2$, we find
\bea
=\frac{1}{(2\pi\epsilon)^d\sqrt{det(V)}} S_{d-2}\Gamma(d) \frac{\pi}{(1+b^T V^{-1} b)^{3/2}}
\\=\frac{1}{2\pi \epsilon^2\sqrt{det(V)}(1+b^T V^{-1} b)^{3/2}},
\eea
and when $d=3$, we find
\bea
=\frac{1}{(2\pi\epsilon)^d} S_{d-2}\Gamma(d) \frac{2}{\sqrt{det(V)}(1+b^T V^{-1} b)^{2}}
\\=\frac{1}{\pi^2\epsilon^3\sqrt{det(V)}(1+b^T V^{-1} b)^{2}}
\eea

When a magnetic field is coupled to $(3+1)$-D edge, as seen in Eq. (20) in the main text, the divergent part of the unregularized free energy is given as,
\bea
F_A=\sum_{n_z<0}2\pi \frac{N_\phi\beta}{L_z}(n_z+\lambda_3)+\sum_{n_z=-\infty,n=1}^\infty 2\pi \frac{N_\phi\beta}{L_z}\sqrt{(n_z+\lambda_3)^2+\alpha n}.
\eea
where $\alpha=B_{field}(\frac{L_z}{2\pi})^2$, $B_{field}$ is the magnetic field and $N_\phi$ is the Landau level degeneracy.
We again write the expression of $\sum_{\vec{m}}e^{-f_A(\vec{m}) \epsilon}$ as done in the case of the free fermions in $(2+1)$-D and $(3+1)$-D to calculate the divergent component. The first term results in the exponential expression
\bea
S_1=\sum_{n_z<0}e^{-2\pi\epsilon \frac{N_\phi\beta}{L_z}(n_z+\lambda_3)},
\eea
while the second term gives
\bea
S_2=\sum_{n_z,n}e^{-2\pi\epsilon\frac{N_\phi\beta}{L_z}\sqrt{(n_z+\lambda_3)^2+\alpha n}}.
\eea
We change both $S_1$ and $S_2$ to integral forms and get the expressions of the poles.
\bea
Integral_1=\int^{-1}_{-\infty} dx e^{-2\pi\epsilon \frac{N_\phi\beta}{L_z}(x+\lambda_3)}
=-\frac{1}{2\pi\epsilon \frac{ N_\phi\beta}{L_z}}e^{-2\pi\epsilon \frac{ N_\phi\beta}{L_z}(\lambda_3-1)},
\eea
\bea
Integral_2=\int^\infty_{-\infty}dx \int^\infty_{0} dy e^{-2\pi \epsilon \frac{N_\phi\beta}{L_z}\sqrt{x^2+\alpha y}}
=\frac{8}{\alpha(2\pi\epsilon N_\phi\frac{\beta}{L_z})^3}.
\eea
\setcounter{equation}{0}
\renewcommand{\theequation}{E\arabic{equation}}
\section{Appendix E: Regularization of gapped Landau level}
Starting from the partition function of Eq. (20) in the main draft, we find the divergent part of the free energy of gapped Landau levels is given by
\bea
F_{A,D}=\sum_{n_3=-\infty}^\infty \sum_{n=0}^\infty(Bn+B+(\frac{2\pi(n_3+\lambda_3)}{L_z})^2)^{1/2}.
\eea
By using the Hurwitz-Zeta function, defined by
\bea
\zeta_H(\epsilon,a)\equiv\sum_{n=0}^\infty (n+a)^{-\epsilon}.
\eea
We can rewrite the above divergent contribution to the partition function as,
\bea
F_{A,Dirac}=\sum_{n_3=-\infty}^\infty \sqrt{B}\zeta_H(-1/2,{(2\pi(n_3+\lambda_3)/L_z)^2}/B+1).
\eea
Using the following Hurwitz-Zeta function identity\cite{Elizalde},
\bea
\zeta_H(\epsilon,x+y)=\sum_{k=0}^{\infty}\frac{\Gamma(\epsilon+k)}{\Gamma(\epsilon)\Gamma(k+1)}\zeta_H(\epsilon+k,x)(-y)^k,
\eea
We may represent the divergent contribution as
\begin{gather}
\zeta_H(-1/2,1+\frac{(2\pi(n_3+\lambda_3)/L_z)^2}{{B}})
\\
\nonumber
=\sum_{k=0}^{\infty}\frac{\Gamma(k-1/2)}{\Gamma(-1/2)\Gamma(k+1)}\zeta(-1/2+k,1)(-\frac{(2\pi(n_3+\lambda_3)/L_z)^2}{{B}})^{k}.
\end{gather}
Then, we rewrite the divergent part of the partition function of the gapped Landau levels as,
\bea
F_{A,D}=\sum_{n_3=-\infty}^\infty \sum_{k=0}^{\infty}  \frac{B^{1/2-k}\Gamma(k-1/2)}{\Gamma(-1/2)\Gamma(k+1)}\zeta(-1/2+k,1)(-(\frac{2\pi}{L_z})^2)^k (n_3+\lambda_3)^{2k}
\\
\nonumber
=\sum_{k=0}^{\infty}  \frac{B^{1/2-k}\Gamma(k-1/2)}{\Gamma(-1/2)\Gamma(k+1)}\zeta(-1/2+k,1)(-(\frac{2\pi}{L_z})^2)^k [\sum_{n_3=-\infty}^\infty  (n_3+\lambda_3)^{2k}]
\eea
From Eq. \eqref{Eq:sumc}, we find that
\bea
\sum_{n_3=-\infty}^\infty  (n_3+\lambda_3)^{2k}=G_1({-k,\lambda_3})=E_1(-k,\lambda_3)+E_1(-k,1-\lambda_3)
\\
\nonumber
=\zeta_H(-2k,\lambda_3)+\zeta_H(-2k,1-\lambda_3)=-\frac{B_{2k+1}(\lambda_3)+B_{2k+1}(1-\lambda_3)}{2k+1}=0.
\eea
Therefore, we conclude that the divergent part of the partition function is regularized to be zero.
\setcounter{equation}{0}
\renewcommand{\theequation}{F\arabic{equation}}
\section{Appendix F: Symmetry Projection}
We consider $N_{edge}$ copies of the Weyl fermions with positive monopole and negative monopole with the magnetic field. When the Weyl fermions with each monopole conserve the fermion number parity. We can consider the partition function of a sector labeled with a definite fermion number parity. This is accomplished by projecting the Hilbert space with the symmetry projection operator, $P$. the symmetry projection operator is given as
\bea
P=\frac{(1+(-1)^{N_{\uparrow}})(1+(-1)^{N_{\downarrow}})}{4}
\eea
where $N_{\uparrow(\downarrow)}$ refers the fermion number operator of positive(negative) monopole. $P$ returns 1 only if $N_{\uparrow}$ and $N_{\downarrow}$ are even integers, thus $P$ projects the Hilbert space into one of the sectors with a definite fermion number parity. Now the partition function with a definite fermion number parity can be expressed as
\begin{gather}
Z_{total,\lambda_3}=Tr(P e^{i \tau H_{\uparrow}}e^{i \tau H_{\downarrow}})
\\
\nonumber
=\frac{1}{4}Tr[(1+e^{ \pi i N_{\uparrow}}) e^{ i \tau H_{\uparrow}}]Tr[(1+e^{\pi i N_{\downarrow}})e^{ i \tau H_{\downarrow}}]
=\frac{Z_{\uparrow\lambda_3}Z_{\downarrow\lambda_3}}{4}.
\end{gather}
where $Z_{\uparrow \lambda_3}$ is given as
\begin{gather}
Z_{\uparrow,\lambda_3}\equiv Tr[(1+e^{ \pi i N_{\uparrow}}) e^{ i \tau H_{\uparrow}}]=[Z_{\lambda_0=0,\lambda_3}]^{N_{edge}}+[Z_{\lambda_0=1/2,\lambda_3}]^{N_{edge}},
\end{gather}
where $H_{\uparrow}^j$ refers the Hamiltonian of the edge with $j$-th flavor and we have used the fact that $Z_{\lambda_0,\lambda_3}=Tr(e^{-2\pi i N \lambda_0}e^{ i \tau H_{\lambda_3}})$\cite{Ryu2012}. We consider the general case of the partition function, which is a linear combination of the periodic, $Z_{\lambda_3=0}$, and anti-periodic, $Z_{\lambda_3=1/2}$, boundary conditions, namely
\bea
Z_{\uparrow}=\sum_{\lambda_0=0,1/2,\lambda_3=0} [Z_{\lambda_0,\lambda_3}]^{N_{edge}}+ s\sum_{\lambda_0=0,1/2,\lambda_3=1/2} [Z_{\lambda_0,\lambda_3}]^{N_{edge}},
\eea
where $s$ is the relative phase between the periodic and antiperiodic sector. From Eq. (21) from the main text, we find that the partition function under the application of the $T$-transformation to be,
\bea
T[Z]=e^{iN_\phi N_{edge}\pi/6}\sum_{\lambda_0=0,1/2,\lambda_3=0}[ Z_{T[\vec{\lambda}]}]^{N_{edge}}+se^{-i\pi N_\phi N_{edge}/12}\sum_{\lambda_0=0,1/2,\lambda_3=1/2} [Z_{T[\vec{\lambda}]}]^{N_{edge}}
\\=e^{iN_\phi N_{edge}\pi /6}[\sum_{\lambda_0=0,1/2,\lambda_3=0}[ Z_{T[\vec{\lambda}]}]]^{N_{edge}}+se^{-i\pi N_\phi N_{edge}/4}\sum_{\lambda_0=0,1/2,\lambda_3=1/2}  [Z_{T[\vec{\lambda}]}]^{N_{edge}}].
\nonumber
\eea
Similarly, under the $S$ transformation, we find
\bea
S[Z]=\sum_{(\lambda_0,\lambda_3)=(0,1/2),(1/2,1/2)}[ s Z_{S[\vec{\lambda}]}]^{N_{edge}}
\\
\nonumber
+e^{i\pi N_\phi {N_{edge}}/2} [Z_{S[\vec{\lambda}=(0,0)]}]^{N_{edge}}+[Z_{S[\vec{\lambda}=(1/2,0)]}]^{N_{edge}}.
\eea

%When $\epsilon_{\mu,\lambda_3}=\epsilon_{\mu,\lambda_3}$, $T[Z]=e^{\pi i}Z$.
This allows us to conclude, as we have done in the main text, that to achieve modular covariance($T[Z]=e^{iN_\phi N_{edge}\pi/6}Z$ and $S[Z]=Z$) $N_\phi N_{edge}$ must be multiples of 8. Therefore, the modular covariance is achieved when $N_{edge}=8/gcd(N_\phi,8)$. When $Z$ has modular covariance, $Z_{total}$ has modular invariance.

%Now we consider a couple of chiral and anti-chiral edges. Suppose we have $\mathbb{Z}_2 \times \mathbb{Z}_2$. Then the partition function is a product of chiral part and anti chiral part
%\bea
%Z=Z_{c}Z_{ac}
%\eea
%Again $S$ is trivial. Under $T$ transformation,
%\bea
%T[Z]=[e^{iN_\phi N\pi 1/6}\sum_{\mu,\lambda_1,\lambda_2=0,1/2,\lambda_3=0}[ Z_{c,T[\vec{\lambda}}]]^N+e^{-i\pi N_\phi N/12}\sum_{\mu,\lambda_1,\lambda_2=0,1/2,\lambda_3=1/2} [Z_{c,T[\vec{\lambda}}]]^N][e^{-iN_\phi N\pi 1/6}\sum_{\mu,\lambda_1,\lambda_2=0,1/2,\lambda_3=0}[ Z_{ac,T[\vec{\lambda}}]]^N+e^{+i\pi N_\phi N/12}\sum_{\mu,\lambda_1,\lambda_2=0,1/2,\lambda_3=1/2} [Z_{ac,T[\vec{\lambda}}]]^N]
%\eea

\end{document}